\documentclass[prb,aps,amssymb,showpacs,twocolumn,superscriptaddress,10pt]{revtex4}
\usepackage{amsmath}
\usepackage{amssymb}
\usepackage{amsthm}
\usepackage{amsfonts}
\usepackage{listings}
\usepackage{enumerate}
\usepackage{latexsym}

\usepackage{psfrag}

\usepackage{bm}
\usepackage[pdftex]{graphicx}
\usepackage{subfigure}

\newcommand{\beq}{\begin{equation}}
\newcommand{\eneq}{\end{equation}}

\newcommand{\ket}[1]{\left|#1\right\rangle}

\def\be{\begin{equation}}
\def\ee{\end{equation}}
\def\ba{\begin{eqnarray}}
\def\ea{\end{eqnarray}}

\def\ie{{\it i.e.},\ }

\input{epsf}

\begin{document}

\tolerance 10000

\newcommand{\vk}{{\bf k}}

\title{Real-Space Entanglement Spectrum of Quantum Hall States}

\author{A. Sterdyniak}
\affiliation{Laboratoire Pierre Aigrain, ENS and CNRS, 24 rue Lhomond, 75005 Paris, France}
\author{A. Chandran}
\affiliation{Department of Physics, Princeton University, Princeton, New Jersey 08544, USA}
\author{N. Regnault}
\affiliation{Laboratoire Pierre Aigrain, ENS and CNRS, 24 rue Lhomond, 75005 Paris, France}
\author{B. A. Bernevig}
\affiliation{Department of Physics, Princeton University, Princeton, New Jersey 08544, USA}
\author{Parsa Bonderson}
\affiliation{Station Q, Microsoft Research, Santa Barbara, California 93106-6105, USA}

\begin{abstract}
We investigate the entanglement spectra arising from sharp real-space partitions of the system for quantum Hall states. These partitions differ from the previously utilized orbital and particle partitions and reveal complementary aspects of the physics of these topologically ordered systems. We show, by constructing one to one maps to the particle partition entanglement spectra, that the counting of the real-space entanglement spectra levels for different particle number sectors versus their angular momentum along the spatial partition boundary is equal to the counting of states for the system with a number of (unpinned) bulk quasiholes excitations corresponding to the same particle and flux numbers. This proves that, for an ideal model state described by a conformal field theory, the real-space entanglement spectra level counting is bounded by the counting of the conformal field theory edge modes. This bound is known to be saturated in the thermodynamic limit (and at finite sizes for certain states). Numerically analyzing several ideal model states, we find that the real-space entanglement spectra indeed display the edge modes dispersion relations expected from their corresponding conformal field theories. We also numerically find that the real-space entanglement spectra of Coulomb interaction ground states exhibit a series of branches, which we relate to the model state and (above an entanglement gap) to its quasiparticle-quasihole excitations. We also numerically compute the entanglement entropy for the $\nu=1$ integer quantum Hall state with real-space partitions and compare against the analytic prediction. We find that the entanglement entropy indeed scales linearly with the boundary length for large enough systems, but that the attainable system sizes are still too small to provide a reliable extraction of the sub-leading topological entanglement entropy term.
\end{abstract}

\date{\today}

\maketitle

\section{Introduction}

Quantum information theory provides powerful tools and perspectives for understanding and characterizing quantum many-body systems. This is particularly true for topologically ordered phases of matter. Such systems possess profoundly entangled states and cannot be characterized by local order parameters, so it is natural to analyze and characterize them through their entanglement properties. One of the most basic (quantum information theoretic) tools for studying entanglement is entropy. If a system in a pure state $\rho$ is partitioned into two subsystems $A$ and $B$, then their entanglement entropy, i.e. the von Neumann entropy
\begin{equation}
S_A = S_B = - \text{Tr} \left[ \rho_A \log \rho_A \right] = - \text{Tr} \left[ \rho_B \log \rho_B \right]
\end{equation}
of either reduced density matrix $\rho_A = \text{Tr}_{B} \left[ \rho \right]$ or $\rho_B = \text{Tr}_{A} \left[ \rho \right]$, is the unique measure of entanglement between the two subsystems.

For a real-space partition of a gapped, two-dimensional (2D) system, the leading order contribution to the entanglement entropy obeys a ``perimeter law,'' i.e., it is proportional to the length $\mathcal{L}$ of the boundary, with a non-universal coefficient $\alpha$, as $\mathcal{L} \rightarrow \infty$~\cite{Bombelli:1986aa,Srednicki:1993aa,Eisert:2010ab}. Kitaev and Preskill~\cite{Kitaev-06prl110404} and Levin and Wen~\cite{levin-06prl110405} showed that the sub-leading, constant term of the entanglement entropy for such a system in its ground state is a universal quantity, which they called the ``topological entanglement entropy.'' In particular, the entanglement entropy (for the ground state) takes the form $S_A = \alpha \mathcal{L} - n \gamma $, up to terms that vanish in the limit $\mathcal{L} \rightarrow \infty$, where $n$ is the number of connected components of the boundary of $A$. Here, $\gamma = \log \mathcal{D}$, where $\mathcal{D} \geq 1$ is a quantity known as the ``total quantum dimension,'' which is characteristic of the system's topological order and equal to $1$ only for trivial topological order (i.e., a gapped system with no anyonic excitations)~\footnote{The total quantum dimension $\mathcal{D} \equiv \sqrt{\sum_{a} d_a^2}$, where $d_a$ are the quantum dimensions of the anyonic charge types that can exist in the corresponding TQFT that describes the long-distance physics of the topologically ordered system. It is also related to the dimension of the ground-state subspace $\mathcal{H}_g$ for the system on a genus $g$ surface by $\text{dim} \mathcal{H}_g = C_g \mathcal{D}^{2g-2}$, with $1 \leq C_g \leq \mathcal{D}^2$ for all $g$. For a $\mathbb{Z}_2$-graded TQFT, i.e. a topological spin theory, which are relevant for the description of fermionic quantum Hall states, the anyonic charges come in $\mathbb{Z}_2$ doublets (e.g. the vacuum and the electron charges form a doublet), so one should use $\tilde{\mathcal{D}} = \mathcal{D}/\sqrt{2}$ as the total quantum dimension of the $\mathbb{Z}_2$-graded theory, in order to count each $\mathbb{Z}_2$ anyonic charge doublet only once.}.

Quantum Hall systems are, so far, the most studied and only experimentally realized topologically ordered phases. Most observed Hall plateaus are expected to be Abelian quantum Hall states~\cite{laughlin83prl1395,haldane83prl605,Halperin83,halperin84prl1583,jain89prl199} with quasiparticles that have fractional charge and exchange (braiding) statistics~\cite{arovas-84prl722}. However, some of the observed plateaus in the second Landau~\cite{Willet-87prl1776,Pan99,Xia04,Kumar2010}, most notably: $\nu=5/2$ and $12/5$, are expected to host non-Abelian states~\cite{Moore-91npb362,Lee:2007,Levin:2007aa,Read-99prb8084,Bonderson:2008aa}, which possess quasiparticles with non-Abelian braiding statistics~\cite{Bonderson:2011aa}. These non-Abelian states have recently received much interest due to their potential application for topologically-protected fault-tolerant quantum computation~\cite{Kitaev03a,Freedman98,Kitaev03b,Nayak:2008aa}.

The entanglement entropy depends on the way the system has been partitioned. Until now, two kinds of bipartite partitions have been applied to fractional quantum Hall (FQH) states: the orbital (or momentum) partition~\cite{haque-07prl060401}, for which $A$ consists of the first $l_A$ orbitals (on the sphere or torus) while $B$ consists of the remaining ones and the particle partition~\cite{PhysRevB.76.125310}, for which $A$ consists of the first $N_A$ particles and $B$ consists of the $N-N_A$ remaining ones. A third partition, the real-space partition, for which $A$ is a spatial region and $B$ is the complement of $A$, has been applied to the integer quantum Hall (IQH) states~\cite{PhysRevB.80.153303}, which can be treated analytically. These different partitions provide access to different properties of the state.
Using an orbital partition as an approximation of a real-space partition (which was argued to be a reasonable approximation, since the orbitals are Gaussian localized), Haque~\textit{et al.}~\cite{haque-07prl060401} have numerically extracted the expected value $\gamma = \frac{1}{2} \ln 3$ for the $\nu=1/3$ Laughlin state. However, the extraction of the topological entanglement entropy (using the orbital partition) in numerical studies has proven to be far less reliable for anything but the simplest model wavefunctions, in particular for the Coulomb Hamiltonian ground states~\cite{zozulya-08prb347}.

Even though the topological entanglement entropy can provide significant information regarding the topological order of a system, it is not able to uniquely identify it. In Ref.~\onlinecite{Kitaev-06prl110404}, Kitaev and Preskill provided an additional heuristic derivation of the topological entanglement entropy, in which they suggested that the reduced density matrix $\rho_A$ for a disk-like region $A$ can be used to \emph{define} a Hamiltonian
\begin{equation}
\mathcal{H}_{A} \equiv - \log \rho_{A}
\end{equation}
which, for a topologically ordered system, would be natural to regard as the Hamiltonian of a $(1+1)$-dimensional conformal field theory (CFT) corresponding to the edge of the system, at least for the purposes of determining the universal contribution to the entanglement entropy. In an effort to develop a more discriminating tool for identifying the topological order of a system (particularly for non-Abelian FQH states), Li and Haldane~\cite{li-08prl010504} proposed examining the (low-lying portion of the) spectrum of the fictitious Hamiltonian $\mathcal{H}_{A} \equiv - \log \rho_{A}$, which they dubbed the ``entanglement spectrum'' (ES), versus the quantum numbers of the available symmetries in the problem.

For both the Moore-Read (MR) model state~\cite{Moore-91npb362} and the ground state of the $\nu=5/2$ Coulomb interaction, Li and Haldane~\cite{li-08prl010504} plotted the ES versus $L_A^z$, the quantum number of angular momentum along the $z$-axis for subsystem $A$. For computational expediency, they used the orbital partition, assuming that, since the orbital basis states are Gaussian localized, this would give a good approximation to the real-space partition. They found that the ES matched the expected MR CFT [i.e., Ising x U(1)] edge mode counting at the lowest pseudo-energies. Given this numerical observation, they suggested (along the lines of Kitaev and Preskill's statement) that the counting of the low-lying ES levels would match the CFT edge mode counting for all quantum Hall states (up to some limit, which grows with the system size). This observation is even more remarkable if one realizes that for generic states the number of entanglement levels should saturate the Hilbert space dimension, a number exponentially larger than the number of CFT edge modes. Renormalization group arguments supporting this conjecture have been given in Ref.~\onlinecite{2011arXiv1103.5437Q}, whereas Ref.~\onlinecite{2011arXiv1102.2218C} proved that the number of levels of the orbital partition ES is bounded from above by the number of the CFT edge modes, thereby proving a large part of the conjecture. In the absence of accidental reductions of the rank of the density matrix, this bound should be saturated.

\begin{figure}[t!]
\includegraphics[width=0.49\linewidth]{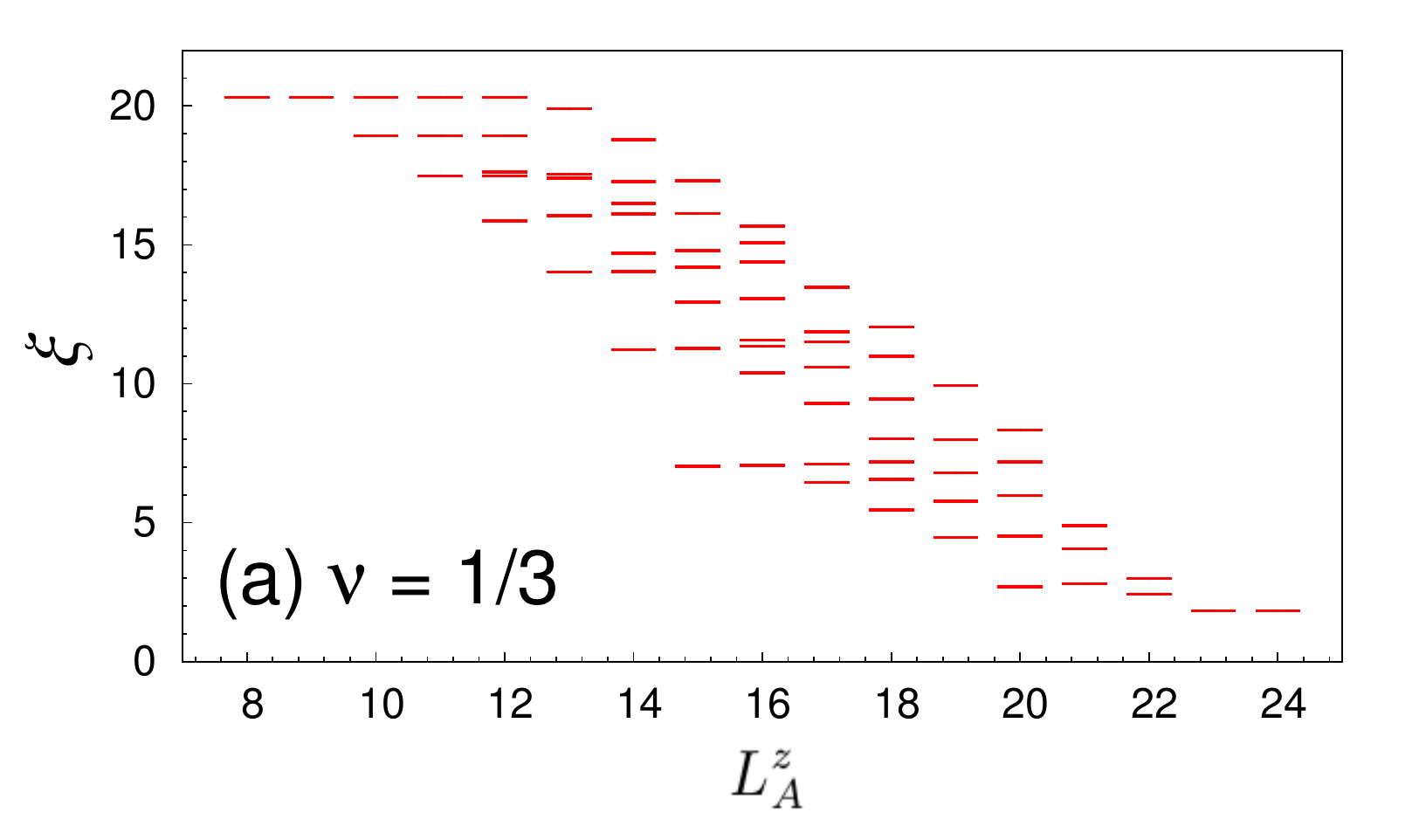}
\includegraphics[width=0.49\linewidth]{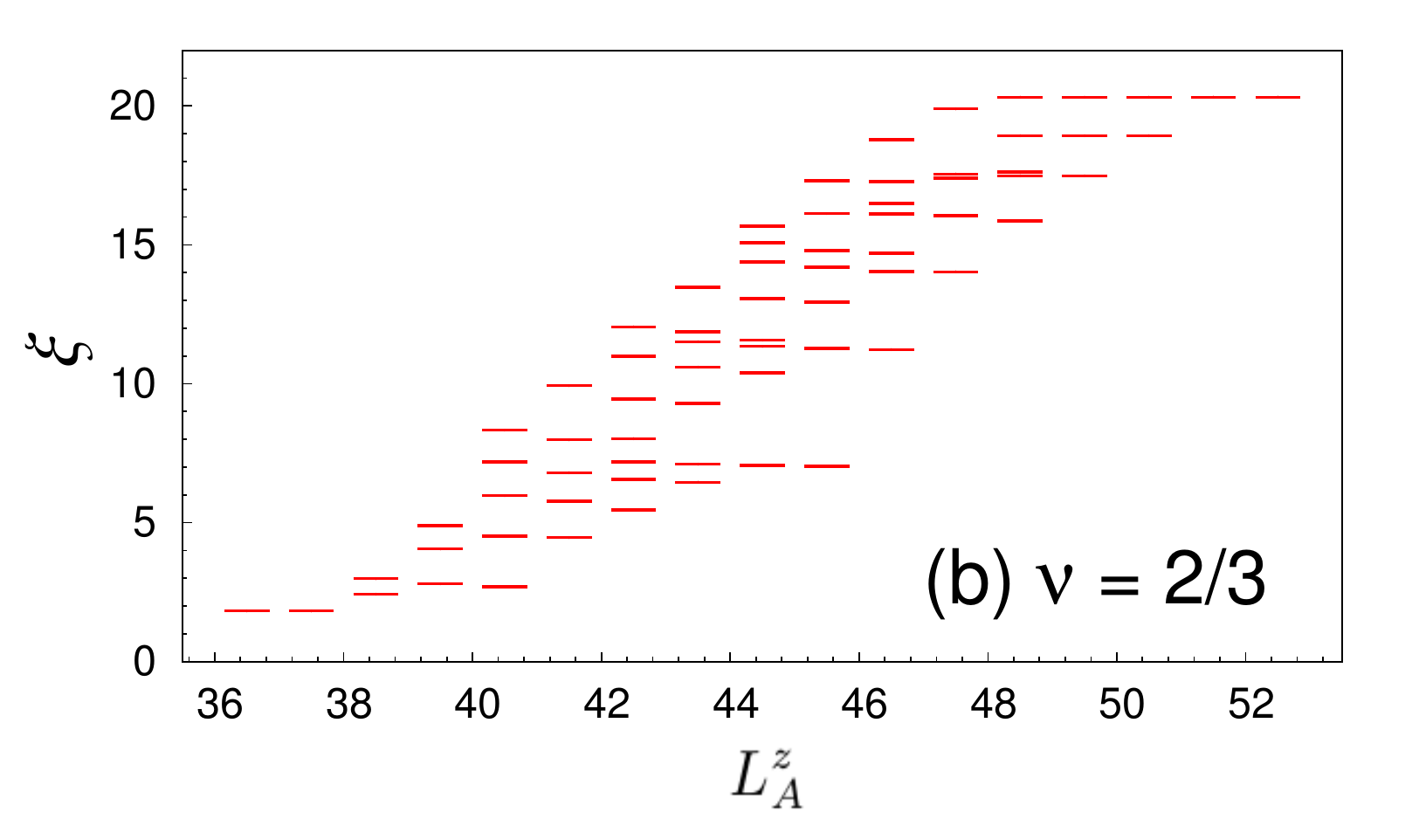}
\caption{(a) The OES for the $\nu=1/3$ Laughlin state with $N=8$ and $N_A=4$. (b) The OES for the $\nu=2/3$ particle-hole conjugate of the $\nu=1/3$ Laughlin state with $N=14$ and $N_A=7$. The two spectra are mirror images of each other: the $\nu= 1/3$ spectrum contains a chiral mode, moving upward toward the left, while the $\nu=2/3$ spectrum contains an anti-chiral mode, moving upward toward the right. The counter-propagating boson mode that should appear in the $\nu=2/3$ state's edge mode spectra does not appear in the OES for the same reason the $\nu=1$ integer quantum Hall state exhibits trivial OES, i.e., the OES probes only the ``interacting'' part of the wavefunction. (We note that the degeneracy between states at $L_A^z = 24, 23$ for $\nu=1/3$ spectrum is a detail due to the fact that, for this particular partition, the two states enjoy an $\vec{L}$ symmetry, which is not present for some other partitions nor for the bosonic version of this state.)}
\label{particleholelaughlin}
\end{figure}

However, since the orbitals have non-zero support across the entire sphere (or plane), the orbital partition entanglement spectra (OES) will not precisely match the real-space partition entanglement spectra (RES). A simple example of this mismatch is provided by the $\nu=1$ IQH state, for which the ground-state wavefunction is the fully filled $n=0$ Landau level. An easy inspection reveals that the ground state $\rho$ is a product state in the orbital basis and, hence, $\rho_A$ is a pure state for any orbital bi-partition (where all $N_A$ orbitals in subsystem $A$ are filled) of this system. Consequently, the ES (and, hence, the entanglement entropy) is trivial, consisting of exactly one level, rather than exhibiting the edge CFT. On the other hand, the edge mode counting is expected to be that of a U(1) CFT (i.e., a free chiral boson). The RES has been computed analytically for non-interacting states~\cite{Peschel:2003aa,PhysRevB.80.153303}, which is done by filling up the levels of a single-particle spectrum (an easily obtainable quantity). For such states, the RES was found to agree with the edge CFT, with a linearly scaling entanglement entropy. Additional examples of the mismatch between the OES and RES are provided by particle-hole conjugation~\cite{PhysRevB.29.6012} of quantum Hall states (such as the $\nu=2/3$ state obtained from the $\nu=1/3$ Laughlin state by particle-hole conjugation). The OES of a state is exactly the \emph{mirror image}, up to a constant shift for $L^z_A$, of that of its particle-hole conjugate state, see Fig.~\ref{particleholelaughlin} for the $\nu=1/3$ and $\nu=2/3$ Laughlin state example. The RES is expected (from the edge CFT) to be given by a combination of the mirror modes with an additional mode propagating counter to these. In a sense, the orbital partition only probes the ``interacting'' part of the state and is not sensitive to the presence of the (non-interacting) integer-filled portion of a quantum Hall state.

In this paper, we examine the ES of quantum Hall states using real-space partitions, both on the sphere and on the disk geometry. The paper is organized as follows: In Section~\ref{sec:RSES}, we provide the general formalism for calculating real-space partition reduced density matrices. In Section~\ref{sec:QHs}, we specialize the formalism to (bosonic and fermionic) quantum Hall states. In Section~\ref{sec:RES_PES}, we show that the RES is related to the particle partition entanglement spectrum (PES)~\cite{PhysRevLett.106.100405} through a one to one map. For quantum Hall states described by a CFT, i.e., with trial wavefunctions generated by conformal blocks, this proves that the RES level counting is bounded by the CFT edge mode level counting. Moreover, it proves that this bound is saturated in the thermodynamic limit for such states and at finite sizes for some states. We expect that the bound will also be saturated at finite sizes by all such states. (This is supported by our numerical results.) In Section~\ref{sec:examples}, we numerically analyze several important examples. In particular, we analyze the $\nu=1$ IQH state, the $\nu=1/3$ Laughlin state and its $\nu=2/3$ particle-hole conjugate state, and the $\nu=1$ bosonic MR state, which all have RES that exhibit the structures expected from the corresponding CFTs. We also analyze the RES of the ground-states  of realistic Coulomb Hamiltonians and find that they exhibit the same low-lying levels as the corresponding model states believed to describe their universality classes. The RES of ground states of these Hamiltonians also exhibit several other branches, besides that of the model states', in the higher ES levels. In Section~\ref{sec:higher_levels}, we relate these higher branches to non-Abelian quasielectron-quasihole excitations of the system for the case of the $\nu=5/2$ Coulomb state. In Section~\ref{sec:annulus}, we analyze the RES for the Laughlin model state as we change the geometry of the manifold from the disk to the annulus to show how the edges of the sample influence the RES. In Section~\ref{sec:quasiholes}, we analyze the RES of the MR state with quasiholes. We show that, when a non-Abelian quasihole crosses the partition boundary, the RES exhibits a sharp change corresponding to the change of topological sectors of the CFT edge mode.
Finally, in Section~\ref{sec:entropy}, we compute the entanglement entropy using the spatial partition for the $\nu=1$ IQH state to verify the validity of the entanglement entropy perimeter law and try to extract the topological entanglement entropy. While we find the entanglement entropy scales linearly with the length of the boundary (for long boundary length), but are unable to reliably compute the topological entanglement entropy even for this simple case. This shows that determining the topological entropy directly from numerics without using prior knowledge about the state is not currently possible.

\section{Real-Space Entanglement Spectrum}
\label{sec:RSES}

We consider a pure, $N$ particle state with wavefunction $\Psi$ and a bipartite real-space partition into regions $A$ and $B$. The (normalized) density matrix is
\begin{widetext}
\begin{equation}
\rho ({\bf r}_1,\ldots,{\bf r}_{N};{\bf r'}_1,\ldots,{\bf r'}_{N}) = \frac{\Psi^*({\bf r}_1,\ldots,{\bf r}_{N}) \Psi({\bf r'}_1,\ldots,{\bf r'}_{N})}{\int \prod\limits_{i = 1}^{N} d^2 {\bf r}_{i} \left| \Psi({\bf r}_1,\ldots,{\bf r}_{N}) \right|^2 }
.
\end{equation}
(We allow for unnormalized wavefunctions, but the density matrices and reduced density matrices throughout this paper are properly normalized to have unit trace.) The reduced density matrix $\rho_A$, obtained by tracing out the degrees of freedom in region $B$, will commute with $\hat{N}_A$, the number operator of particles in region $A$, so the density matrix is block diagonal in sectors of different $N_A$, which can be considered separately. In particular,
\begin{equation}
\rho_A = \sum_{N_A = 0}^{N} p_{N_A} \, \rho_{N_{A}} ({\bf r}_1,\dots,{\bf r}_{N_A};{\bf r'}_1,\dots,{\bf r'}_{N_A})
,
\end{equation}
where
\begin{equation}
p_{N_A} = \binom{N}{N_A} \int_A \prod_{i = 1}^{N_A} d^2 {\bf r}_{i} \int_B \prod_{j = N_A+1}^{N} d^2 {\bf r}_{j} \,\, \rho ({\bf r}_1,\ldots,{\bf r}_{N};{\bf r}_1,\ldots,{\bf r}_{N})
\end{equation}
is the probability of the reduced system having $N_A$ electrons in region $A$ with (normalized) density matrix
\begin{eqnarray}
&& \rho_{N_A} ({\bf r}_1,\dots,{\bf r}_{N_A};{\bf r'}_1,\dots,{\bf r'}_{N_A}) = \frac{\int_B \prod\limits_{j = N_A+1}^{N} d^2 {\bf r}_{j} \,\, \rho ({\bf r}_1,\ldots,{\bf r}_{N_A},{\bf r}_{N_A +1},\ldots,{\bf r}_{N};{\bf r'}_1,\ldots,{\bf r'}_{N_A},{\bf r}_{N_A +1},\ldots,{\bf r}_{N}) }
{\int_A \prod\limits_{i = 1}^{N_A} d^2 {\bf r}_{i} \int_B \prod\limits_{j = N_A+1}^{N} d^2 {\bf r}_{j} \,\, \rho ({\bf r}_1,\ldots,{\bf r}_{N};{\bf r}_1,\ldots,{\bf r}_{N})} \notag \\
&& \qquad \qquad = \frac{1}{p_{N_A}} \binom{N}{N_A} \int_B \prod\limits_{j = N_A+1}^{N} d^2 {\bf r}_{j} \,\, \rho ({\bf r}_1,\ldots,{\bf r}_{N_A},{\bf r}_{N_A +1},\ldots,{\bf r}_{N};{\bf r'}_1,\ldots,{\bf r'}_{N_A},{\bf r}_{N_A +1},\ldots,{\bf r}_{N})
\label{realdm}
\end{eqnarray}
where the coordinates of the $N_A$ remaining particles are restricted to region $A$ (and the $N_B = N - N_A$ particles confined to region $B$ have been traced out). We note that if $\rho$ is a product of (identical) single-particle states, then $p_{N_A} = \binom{N}{N_A} p_{A}^{N_A} (1-p_{A})^{N-N_A}$, where $p_A$ is the probability for any one of the particles to be in region $A$, is a binomial distribution (which becomes a Gaussian distribution as $N \rightarrow \infty$). In the rest of the paper, we will work with the density matrix $\rho_{N_A}$ of a specific $N_A$ subspace, and leave the coordinates implicit when the context is clear.

\subsection{Quantum Hall States}
\label{sec:QHs}

A general bosonic or fermionic quantum Hall wavefunction for $N$ particles in the lowest Landau level with total $z$-axis angular momentum $L_{\text{tot}}^z$ along the $z$-axis can be expressed as linear combinations of Fock states of single particle orbitals: $\Psi = \sum_{{\boldsymbol \lambda}} c_{{\boldsymbol \lambda}} \mathcal{M}_{{\boldsymbol \lambda}}$. Here ${\boldsymbol \lambda} = (\lambda_1,\dots,\lambda_{N})$ is a length $N$ partition of $L_{\text{tot}}^z$, \ie $\|{\boldsymbol \lambda} \| \equiv \sum_i \lambda_i = L_{\text{tot}}^z$ and $\lambda_{i} \geq \lambda_{i+1}$. The normalized symmetric monomial (for bosonic systems) or Slater determinant (for fermionic systems) that has orbital occupation corresponding to the partition ${\boldsymbol \lambda}$ is given by
\begin{equation}
\mathcal{M}_{{\boldsymbol \lambda}} = \frac{1}{\sqrt{N ! \prod\limits_{m} \left[ n_{m}({\boldsymbol \lambda}) \right]! }} \sum\limits_{\sigma\in\mathcal{S}_N} \epsilon(\sigma) \prod\limits_{i=1}^{N} \phi_{\lambda_{\sigma(i)}}({\bf r}_{i}) ,
\end{equation}
where $\mathcal{S}_N$ is the permutation group of $N$ objects, $\epsilon(\sigma) = (\pm 1)^{\sigma}$ is equal to $1$ for bosonic systems or the signature of the permutation $\sigma$ for fermionic systems, $n_{m}({\boldsymbol \lambda})$ is the number of occurrences of the integer $m$ in the partition ${\boldsymbol \lambda}$, corresponding to an occupation of the $m$th orbital (which are all equal to $0$ or $1$ for the fermionic case), and
\begin{equation}
\label{onebodywf}
\phi_m ({\bf r})= \left\{
\begin{array}{lcc}
\frac{1}{\sqrt{2 \pi  2^{m} m! }} z^m e^{-\frac{1}{4} |z|^2} & & \text{ plane} \\
& & \\
\sqrt{\frac{(N_{\phi}+1)! }{4 \pi m! (N_{\phi} - m)! } } \left[ \cos(\theta/2) e^{i \varphi/2} \right]^{m} \left[ \sin(\theta/2) e^{-i \varphi/2} \right]^{ N_{\phi} - m} & & \text{ sphere}
\end{array}
\right.
\end{equation}
is the single particle orbital with $L^z$ eigenvalue $m$ on the plane with coordinates $z=x+iy$ (where $m=0,1,\ldots$), or with $L^z$ eigenvalue $m - N_{\phi}/2$ of the sphere with polar coordinates $(\theta,\varphi)$ and $N_{\phi}$ flux quanta through the surface of the sphere (where $m=0,1,\ldots,N_{\phi}$). The functions $\mathcal{M}_{{\boldsymbol \lambda}} $ form a set of orthonormal free many-body states.

Using Eq.~(\ref{realdm}), we can derive the expression of the reduced density matrix for the real space partition as a function of the coefficients $c_{{\boldsymbol \lambda}}$:
\begin{eqnarray}
\rho_{N_A}  &=& \frac{1}{ p_{N_A} } \binom{N}{N_A} \sum_{{\boldsymbol \lambda},{\boldsymbol \lambda'}} c^*_{{\boldsymbol \lambda}}c_{{\boldsymbol \lambda'}} \prod_{j = N_A+1}^N  \int_B d^2 {\bf r}_j \mathcal{M}^*_{{\boldsymbol \lambda}} \mathcal{M}_{{\boldsymbol \lambda'}} \notag \\
&=& \frac{1}{ p_{N_A} N_A ! N_B ! } \sum_{{\boldsymbol \lambda},{\boldsymbol \lambda'}}  \frac{c^*_{{\boldsymbol \lambda}}c_{{\boldsymbol \lambda'}}}{ \sqrt{ \prod\limits_{m} \left[ n_{m}({\boldsymbol \lambda}) \right]! \left[ n_{m}({\boldsymbol \lambda'}) \right]! }} \sum_{\sigma,\tau \in \mathcal{S}_N}  \epsilon(\sigma) \epsilon(\tau)\prod_{i=1}^{N_A} \phi^*_{\lambda_{\sigma(i)}}({\bf r}_{i}) \phi_{\lambda'_{\tau(i)}}({\bf r}'_{i}) \notag \\
&& \qquad \qquad \qquad \qquad \qquad \qquad \qquad \qquad \times \prod_{j = N_A+1}^N  \int_B d^2 {\bf r}_j \phi^*_{\lambda_{\sigma(j)}}({\bf r}_{j}) \phi_{\lambda'_{\tau(j)}}({\bf r}_{j})
.
\label{realdm2}
\end{eqnarray}

To proceed further, we need to specify the spatial regions $A$ and $B$. While a generic integration domain can certainly be used, we want to minimize our computational burden and also be able to compare the resulting ES with that of previous approaches. For these reasons, we chose to use domains that are rotationally invariant around the $z$-axis, which preserves the $L^z$ symmetry. The projection on the $z$-axis of the angular momentum for the particles in region $A$, $L^z_A$, was key in identifying the model state structure within the ES of realistic system~\cite{li-08prl010504}. On the plane, we take region $A$ to be the disk-shaped domain centered at the origin with radius $R$ (i.e. the region $r<R$). On the sphere, we take $A$ to be the cap centered at the north pole up to azimuthal angle $\Theta$ (i.e., the region $\theta <\Theta$). For such domains, the integrals of Eq.~(\ref{realdm2}) will lead to a vanishing contribution unless $\lambda_{\sigma(j)} = \lambda'_{\tau(j)}$ for all $j = N_A+1,\ldots,N$. This ensures that $[\rho_A,\hat{L}^z_A] =0$ and, thus, $\rho_{A}$ is also block-diagonal in $L_A^z$ sectors. To write $\rho_{N_A}$ more explicitly, we use the property
\begin{equation}
\label{eq:M_decomp}
\mathcal{M}_{{\boldsymbol \lambda}} ({\bf r}_1,\ldots,{\bf r}_{N})= \sum_{\substack{{\boldsymbol \mu},{\boldsymbol \nu} \\  \left\langle {\boldsymbol \mu}; {\boldsymbol \nu} \right\rangle = {\boldsymbol \lambda} }} \epsilon({\boldsymbol \mu},{\boldsymbol \nu}) \left( \frac{N_A ! N_B!}{N!} \prod\limits_{m} \frac{ \left[ n_{m}({\boldsymbol \lambda}) \right]! } { \left[ n_{m}({\boldsymbol \mu}) \right] ! \left[ n_{m}({\boldsymbol \nu}) \right]!} \right)^{\frac{1}{2} } \, \mathcal{M}_{{\boldsymbol \mu}} ({\bf r}_1,\ldots,{\bf r}_{N_A}) \mathcal{M}_{{\boldsymbol \nu}} ({\bf r}_{N_A+1},\ldots,{\bf r}_{N})
\end{equation}
where ${\boldsymbol \mu}$ are length $N_A$ partitions, ${\boldsymbol \nu}$ are length $N_B$ partitions, we define $\left\langle {\boldsymbol \mu}; {\boldsymbol \nu} \right\rangle$ to be the ordered partition with elements $\mu_1, \ldots, \mu_{N_A}, \nu_{1} , \ldots , \nu_{N_B}$, we define $\epsilon({\boldsymbol \mu},{\boldsymbol \nu}) = \epsilon(\sigma)$ for $\sigma \in \mathcal{S}_N$ such that $(\lambda_{\sigma(1)} , \ldots, \lambda_{\sigma(N)}) = ( \mu_1, \ldots, \mu_{N_A}, \nu_{1} , \ldots , \nu_{N_B} )$ for ${\boldsymbol \lambda} = \left\langle {\boldsymbol \mu}; {\boldsymbol \nu} \right\rangle$ (i.e. $\sigma^{-1}$ is the permutation needed to combine ${\boldsymbol \mu}$ and ${\boldsymbol \nu}$ into an ordered, length $N$ partition), and the sum is over all such partitions that can be combined to give the partition ${\boldsymbol \lambda}$ (i.e. ${\boldsymbol \mu}$ and ${\boldsymbol \nu}$ are subjected to the constraint $\left\langle {\boldsymbol \mu}; {\boldsymbol \nu} \right\rangle = {\boldsymbol \lambda}$). We also define
\begin{equation}
\mathcal{F}_B (m) = \int_B d^2 {\bf r} \, |\phi_{m}({\bf r})|^2  =
  \begin{cases}
   Q(m+1,\frac{R^2}{2}) & \text{plane}\\
   &\\
   I_{\cos^2(\frac{\Theta}{2})} (m+1,N_{\phi}-m+1) & \text{sphere}
  \end{cases}
\end{equation}
to be the norm-squared of the orbitals restricted to region $B$, where $Q(a,x) = \Gamma(a,x) / \Gamma(a)$ is the regularized incomplete gamma function and $I_{x}(a,b) = B(x;a,b) / B(a,b)$ is the regularized incomplete beta function. This allows us to write
\begin{eqnarray}
\rho_{N_A} &=& \sum_{{\boldsymbol \mu},{\boldsymbol \mu'};{\boldsymbol \nu}} R^*_{{\boldsymbol \mu},{\boldsymbol \nu}} R_{{\boldsymbol \mu'},{\boldsymbol \nu}} \mathcal{M}^*_{{\boldsymbol \mu}} \mathcal{M}_{{\boldsymbol \mu'}} \\
R_{{\boldsymbol \mu},{\boldsymbol \nu}} &=& \frac{1}{\sqrt{ p_{N_A} }} c_{\left\langle {\boldsymbol \mu}; {\boldsymbol \nu} \right\rangle} \epsilon({\boldsymbol \mu},{\boldsymbol \nu})  \sqrt{ \prod\limits_{m} \frac{ \left[ n_{m}(\left\langle {\boldsymbol \mu}; {\boldsymbol \nu} \right\rangle ) \right]! } { \left[ n_{m}({\boldsymbol \mu}) \right] ! \, \left[ n_{m}({\boldsymbol \nu}) \right]! }} \,\, \sqrt{ \prod_{j = 1}^{N_B} \mathcal{F}_B(\nu_j) }
.
\end{eqnarray}
We note that this indicates that the RES will weight the different orbitals according to the overlaps of orbitals in the traced out region $B$.

We can also write this in terms of $L_A^z$ sectors as
\begin{eqnarray}
\rho_{N_A} &=& \sum_{L_A^{z} = 0}^{L_{\text{tot}}^{z}} p(L_A^z | N_A ) \rho_{N_A, L^z_A} \\
\rho_{N_A, L^z_A} &=& \frac{1}{p(L_A^z | N_A )} \sum_{\substack{ {\boldsymbol \mu},{\boldsymbol \mu'} \\ \|{\boldsymbol \mu}\| = \| {\boldsymbol \mu'} \| = L_{A}^z }} \sum_{\substack{ {\boldsymbol \nu} \\\| {\boldsymbol \nu} \| = L_{B}^z }} R^*_{{\boldsymbol \mu},{\boldsymbol \nu}} R_{{\boldsymbol \mu'},{\boldsymbol \nu}} \mathcal{M}^*_{{\boldsymbol \mu}} \mathcal{M}_{{\boldsymbol \mu'}}
\end{eqnarray}
\end{widetext}
where $p(L_A^z | N_A )$ is the conditional probability of the reduced system $A$ having angular momentum $L_A^z$, given that it has $N_A$ particles, and $p_{N_A,L_A^z} = p(L_A^z | N_A ) p_{N_A}$ is the probability that the reduced system $A$ has $N_A$ particles and angular momentum $L_A^z$.

We emphasize that the coordinates are restricted to the region $A$ and so one might want to write the reduced density matrices in terms of basis states that are normalized on region $A$. Defining
\begin{eqnarray}
\phi^A_m ({\bf r}) &=& \frac{\phi_{m}({\bf r})}{\sqrt{1-\mathcal{F}_B(m)}} \\
\mathcal{M}^{A}_{{\boldsymbol \mu}} &=& \frac{\mathcal{M}_{{\boldsymbol \mu}} }{\sqrt {\prod_{i = 1}^{N_A} \left[1- \mathcal{F}_B(\mu_i) \right] }} \\
R^{A}_{{\boldsymbol \mu},{\boldsymbol \nu}} &=&  \sqrt {\prod_{i = 1}^{N_A} \left[1- \mathcal{F}_B(\mu_i) \right] } \,\, R_{{\boldsymbol \mu},{\boldsymbol \nu}}
\end{eqnarray}
it is clear that we can simply substitute these in for their corresponding quantities in the above equations. Using $\mathcal{M}^{A}_{{\boldsymbol \mu}}$ to define orthonormal basis states $\left| {\boldsymbol \mu}_{A} \right\rangle$ for subsystem $A$, we can thus write the density matrix operator as
\begin{equation}
\hat{\rho}_{N_A} = \sum_{{\boldsymbol \mu},{\boldsymbol \mu'}; {\boldsymbol \nu}} R^{A *}_{{\boldsymbol \mu},{\boldsymbol \nu}} R^{A}_{{\boldsymbol \mu'},{\boldsymbol \nu}} \left| {\boldsymbol \mu'}_{A} \right\rangle \left\langle {\boldsymbol \mu}_{A} \right|
.
\end{equation}
In other words, the density matrix elements in this basis are
\begin{equation}
\label{eq:rho_N_A_elements}
[\hat{\rho}_{N_A}]_{{\boldsymbol \mu'},{\boldsymbol \mu}} = \sum_{{\boldsymbol \mu},{\boldsymbol \mu'}; {\boldsymbol \nu}} R^{A *}_{{\boldsymbol \mu},{\boldsymbol \nu}} R^{A}_{{\boldsymbol \mu'},{\boldsymbol \nu}} = [R^A R^{A \dagger}]_{{\boldsymbol \mu'},{\boldsymbol \mu}}
\end{equation}
where on the right hand side we are treating $R^A$ as a matrix with row indices ${\boldsymbol \mu}$ and column indices ${\boldsymbol \nu}$.

It is worth pointing out a distinction between the orbital and real-space partitions that generally appears in the probability distribution $p_{N_A}$ of having $N_A$ particles in subsystem $A$. As the (non-interacting) IQH states are product states, the probabilities $p_{N_A}$ form a binomial distribution. On the other hand, applying an orbital partition to an IQH state gives $p_{N_A} = \delta_{N_A , N_A^{\text{orb}}}$, where $N_A^{\text{orb}}$ is the total number of orbitals comprising subsystem $A$. More generally, one expects the interacting quantum Hall ground states to exhibit a probability distribution $p_{N_A}$ that is qualitatively similar for the real-space partition, i.e., a nearly binomial distribution peaked around $N_A \approx f_A N$, where $f_A = \frac{\text{area}(A)}{\text{area}(A \cup B)}$. Similarly, the probability distribution for the orbital partition will be sharply peaked around $N_A \approx f_A N$, where here $f_A = \frac{N_A^{\text{orb}}}{N_A^{\text{orb}} + N_B^{\text{orb}}}$, and will drop to \emph{exactly} zero for values of $N_A$ beyond some range away from $f_A N$, as dictated by the squeezing properties of the state.

\subsection{Relation between real-space partition and particle partition entanglement spectra}
\label{sec:RES_PES}

We can now demonstrate that there is a one to one map between the RES and the PES of an arbitrary $N$ particle state $\ket{\Psi}$. For a particle partition of the system into subsystems $\tilde{A}$ and $\tilde{B}$ comprised of $N_A$ and $N_B$ particles, respectively, the reduced density matrix operator for subsystem $\tilde{A}$ is obtained by removing $N_B$ particles, \emph{irrespective} of their position in space. This is given by
\begin{widetext}
\begin{equation}
\rho_{\tilde{A}} ({\bf r}_1,\dots,{\bf r}_{N_A};{\bf r'}_1,\dots,{\bf r'}_{N_A}) = \int \prod\limits_{j = N_A+1}^{N} d^2 {\bf r}_{j} \,\, \rho ({\bf r}_1,\ldots,{\bf r}_{N_A},{\bf r}_{N_A +1},\ldots,{\bf r}_{N};{\bf r'}_1,\ldots,{\bf r'}_{N_A},{\bf r}_{N_A +1},\ldots,{\bf r}_{N})
\label{partdm}
\end{equation}
where the domains of the integrals and the coordinates of the remaining particles are the entire physical space of the system. We notice, apart from normalizations and the difference of domains of the integrations and remaining coordinates, this has a form similar to $\rho_{N_A}$ of Eq.~(\ref{realdm}), so it should be clear that the two can be related.

Using the decomposition of Eq.~(\ref{eq:M_decomp}), we can write the wavefunction $\Psi = \sum_{{\boldsymbol \lambda}} c_{{\boldsymbol \lambda}}\mathcal{M}_{{\boldsymbol \lambda}}$ as
\begin{equation}
\Psi = \sum_{\substack{{\boldsymbol \mu},{\boldsymbol \nu} \\  \left\langle {\boldsymbol \mu}; {\boldsymbol \nu} \right\rangle = {\boldsymbol \lambda} }} c_{\left\langle {\boldsymbol \mu} ; {\boldsymbol \nu} \right\rangle } \epsilon({\boldsymbol \mu},{\boldsymbol \nu}) \left( \frac{N_A ! N_B!}{N!} \prod\limits_{m} \frac{ \left[ n_{m}(\left\langle {\boldsymbol \mu} ; {\boldsymbol \nu} \right\rangle) \right]! } { \left[ n_{m}({\boldsymbol \mu}) \right] ! \left[ n_{m}({\boldsymbol \nu}) \right]!} \right)^{\frac{1}{2} }  \mathcal{M}_{{\boldsymbol \mu}} ({\bf r}_1,\ldots,{\bf r}_{N_A}) \mathcal{M}_{{\boldsymbol \nu}} ({\bf r}_{N_A+1},\ldots,{\bf r}_{N})
.
\label{pem}
\end{equation}
Applying the particle partition, the reduced density matrix operator's elements are given simply by
\begin{equation}
[\hat{\rho}_{\tilde{A}}]_{{\boldsymbol \mu'},{\boldsymbol \mu}} = \sum_{{\boldsymbol \nu}}  c^*_{\left\langle {\boldsymbol \mu} ; {\boldsymbol \nu} \right\rangle } c_{\left\langle {\boldsymbol \mu'} ; {\boldsymbol \nu} \right\rangle } \epsilon({\boldsymbol \mu},{\boldsymbol \nu}) \epsilon({\boldsymbol \mu'},{\boldsymbol \nu}) \frac{N_A ! N_B!}{N!} \left(  \prod\limits_{m} \frac{ \left[ n_{m}(\left\langle {\boldsymbol \mu} ; {\boldsymbol \nu} \right\rangle) \right]! \left[ n_{m}(\left\langle {\boldsymbol \mu'} ; {\boldsymbol \nu} \right\rangle) \right]! } { \left[ n_{m}({\boldsymbol \mu}) \right] ! \left[ n_{m}({\boldsymbol \mu'}) \right] ! \left( \left[ n_{m}({\boldsymbol \nu}) \right]! \right)^2 } \right)^{\frac{1}{2} } = [P P^{\dagger}]_{{\boldsymbol \mu'},{\boldsymbol \mu}}
\end{equation}
where we have defined the matrix $P$ with elements
\begin{equation}
[P]_{{\boldsymbol \mu} , {\boldsymbol \nu} } = c_{\left\langle {\boldsymbol \mu} ; {\boldsymbol \nu} \right\rangle } \epsilon({\boldsymbol \mu},{\boldsymbol \nu}) \sqrt{\frac{N_A ! N_B!}{N!}} \left(  \prod\limits_{m} \frac{ \left[ n_{m}(\left\langle {\boldsymbol \mu} ; {\boldsymbol \nu} \right\rangle) \right]! } { \left[ n_{m}({\boldsymbol \mu}) \right] ! \left[ n_{m}({\boldsymbol \nu}) \right]! } \right)^{\frac{1}{2} }
\end{equation}
\end{widetext}
with row indices ${\boldsymbol \mu}$ and column indices ${\boldsymbol \nu}$.

Comparing this to $\hat{\rho}_{N_A}$ in Eq.~(\ref{eq:rho_N_A_elements}), we can define the diagonal matrices $Q$ and $S$ with elements
\begin{equation}
[Q]_{{\boldsymbol \nu} , {\boldsymbol \nu'} } = \sqrt{ \frac{1}{p_{N_A} } \binom{N}{N_A} \prod_{j=1}^{N_B} \mathcal{F}_B (\nu_j) } \,\, \delta_{{\boldsymbol \nu} , {\boldsymbol \nu'}}
\end{equation}
and
\begin{equation}
[S]_{{\boldsymbol \mu} , {\boldsymbol \mu'} } = \sqrt{ \prod_{i=1}^{N_A} \left[ 1- \mathcal{F}_B (\mu_i) \right] } \,\, \delta_{{\boldsymbol \mu} , {\boldsymbol \mu'}}
\end{equation}
to give us the relation $R^A = S R = S P Q$. Within the many-body basis, the calculations of the RES can now be done numerically in a way similar to the calculations of PES.

Since $Q$ and $S$ are diagonal with no zeros on the diagonal, $R^A$, $R$, and $P$ have the same rank. Consequently, the reduced density matrix operators $\hat{\rho}_{\tilde{A}}$ of the particle partition and $\hat{\rho}_{N_{A}}$ of the real-space partition in the $N_A$ sector have the same rank. It is clear that this also holds when one restricts to a particular $L_A^z$ or $L_{\tilde{A}}^z$ sector. However, it is worth mentioning that, on the sphere, the total angular momentum $L_A^2$ of subsytem $A$ is a good quantum number for the PES, but is not a good quantum number for the RES and the OES. This gives a one to one mapping between the levels of the RES and the PES. For states described by a CFT, the counting of the PES levels per momentum sector was shown to equal the counting of states for the system with unpinned bulk quasiholes corresponding to the same number of particles $N_A$ and orbitals $N_\phi+1$~\cite{PhysRevLett.106.100405}. It has been proven for such quantum Hall states that this counting of states with quasiholes is bounded by the level counting of the corresponding CFT edge modes~\cite{PhysRevLett.106.100405}. Moreover it has been proven that this bound is saturated in the thermodynamic limit for all such quantum Hall states and also at finite sizes for Laughlin states~\cite{Estienne-unpublished}. Thus, we have proven that the RES level counting of such model states similarly satisfies and saturates the bound by the CFT edge mode counting.

\section{Examples}
\label{sec:examples}

\subsection{$\nu=1$ Integer Quantum Hall}

\begin{figure}[t!]
\includegraphics[width=\linewidth]{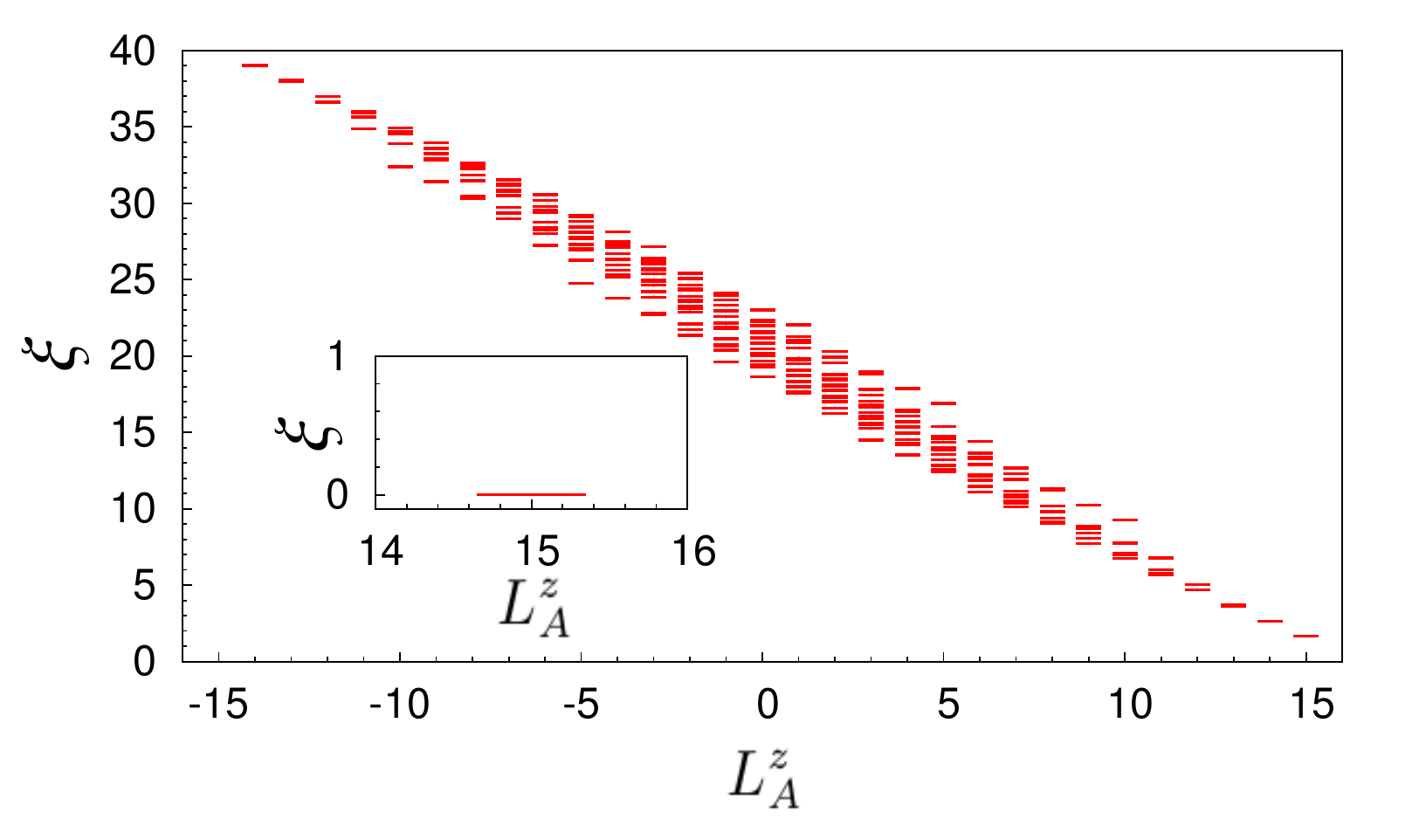}
\caption{The RES of the $\nu = 1$ slater state for $N=11$ particles with $N_A=5$. The sphere was partition into two hemispheres ($\Theta = \pi/2$). Inset: The OES for the same state ($N_A = 5)$.}
\label{realpartitionslater}
\end{figure}

\begin{figure}[t!]
\includegraphics[width = \linewidth]{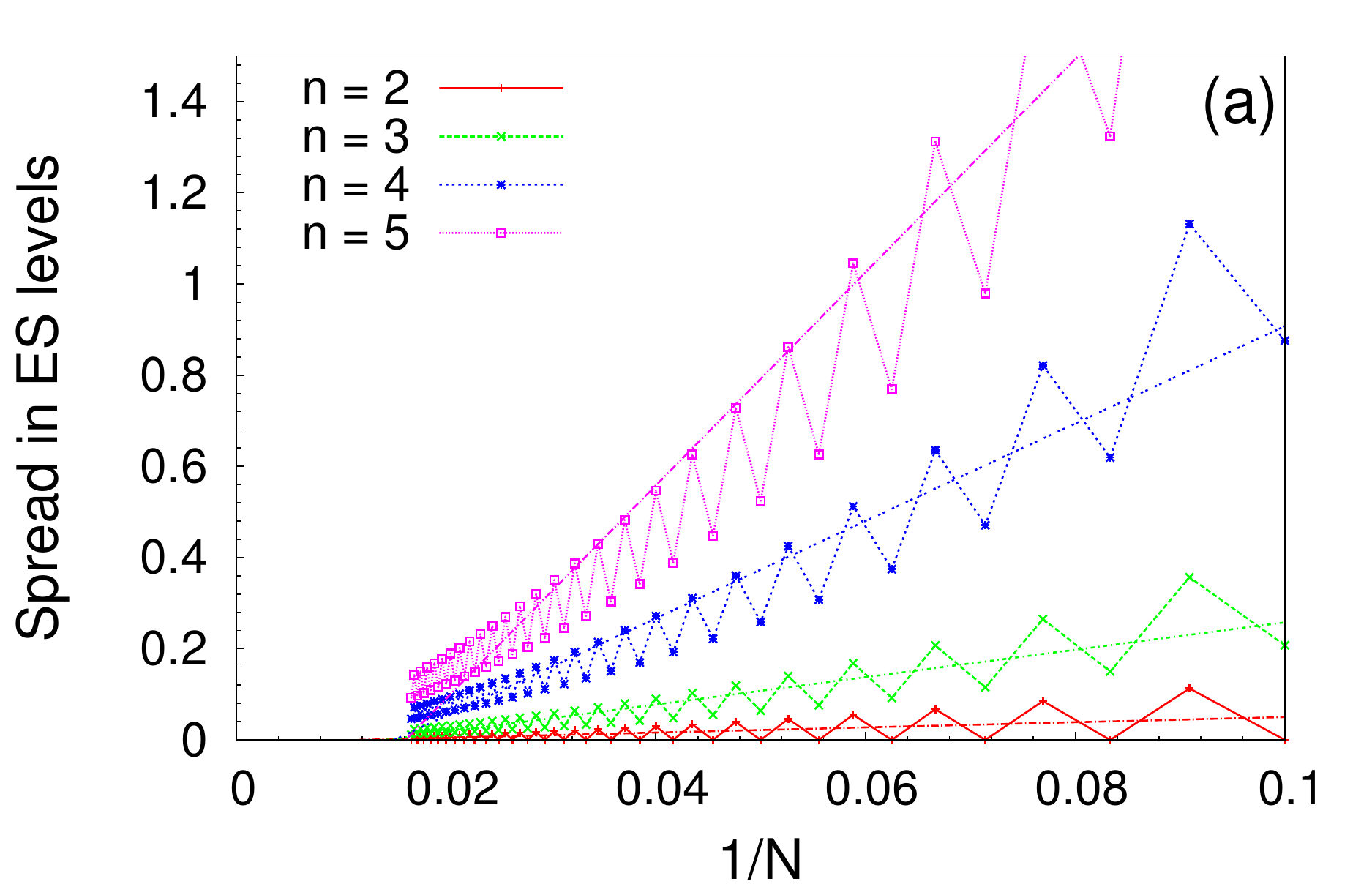}
\includegraphics[width = \linewidth]{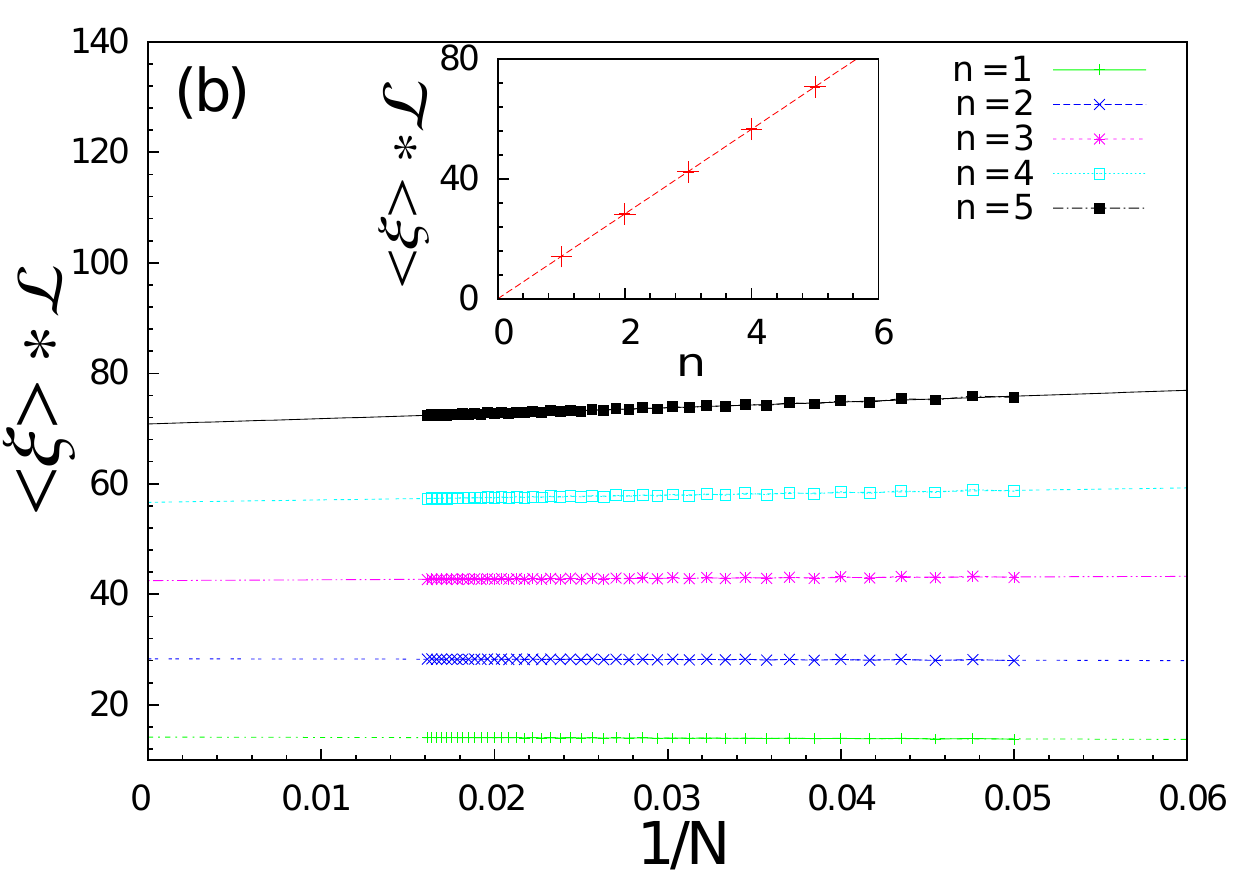}
\caption{(a) Spread in RES levels vs. 1/$N$ for different $L_A^z$ sectors for $\Theta = \pi/2$. An even-odd particle number effect is observed, and the spread of the levels vanishes in the thermodynamic limit. (b) Value of the zero-spread ES-levels [as shown in (a), the spread of the levels for each $L_A^z$ vanishes in the thermodynamic limit, which means that all the entanglement levels at a particular $L_A^z$ collapse to the same value; this value is the zero-spread-ES level] extrapolated at the thermodynamic limit. Here $n=\max[L_A^z]-L_A^z$. The velocity of the edge mode is $ v = 2.254(3)$.}
\label{slateredgedispersion}
\end{figure}

As a first example, we consider the $\nu = 1$ IQH state. In the orbital basis, this state can always be written as a tensor product $\ket{111\dots11} =  \ket{11}_A \otimes \ket{1\dots11}_B$, where $\ket{111\dots11}$ means that each one of the $N_{\phi }+1 = N$ orbitals is occupied by exactly one electron. Thus, for any orbital partition, its OES always consists in a single state (see inset of Fig.~\ref{realpartitionslater}), which does not correspond to its expected edge mode. This shortcoming is averted by the use of the real-space partition. As shown in Fig.~\ref{realpartitionslater}, the RES exhibits a linear dispersion relation. (Note that all our entanglement spectra are displayed for $N_A=N/2$, which is the one that contains the largest amount of information.) As expected, the counting in each $(N_A,L^z_A)$ sector is given by the counting of states for a number unpinned bulk quasiholes corresponding to the same quantum numbers. This equals the number of ways to put $N_A$ particles in $N_{\phi + 1}$ orbitals such that the sum of their orbital numbers equals $L^z_A$. Note that the RES for the $\nu=1$ state exhibits a chiral structure of levels.

The spread of the levels at each $L_A^z$ scales to zero in the thermodynamic limit, as can be seen in Fig.~\ref{slateredgedispersion}a, and the spectrum has a linear shape reminiscent of a (relativistic) CFT edge theory's energy dispersion. As such, we can compute the RES's ``edge velocity,'' which is the overall slope of the entanglement spectrum. The calculation uses the following procedure: We partition the system into two hemispheres ($\Theta=\pi / 2$) and focus on $N_A=N/2$, so that the filling factor in $A$ is identical to that of the original system. We then plot the mean value of $\xi \mathcal{L}$ for a given $L_A^z$ sector as a function of $1/N$. From these values, we subtract the zero momentum energy obtained at the maximal value of $L_A^z$. If RES mimics the edge mode's true energy spectrum in the thermodynamical limit, the extrapolated values of $\xi \mathcal{L}$ should be equal to $2 \pi v (\max [ L_A^z ] - L_A^z)$ when $N\rightarrow\infty$, where $v$ is the edge mode velocity. The extrapolation, shown in Fig.~\ref{slateredgedispersion}b, gives an extracted velocity of $v=2.254(3)$.

\subsection{$\nu=1/3$ Fractional Quantum Hall}

\begin{figure}[t!]
\includegraphics[width=\linewidth]{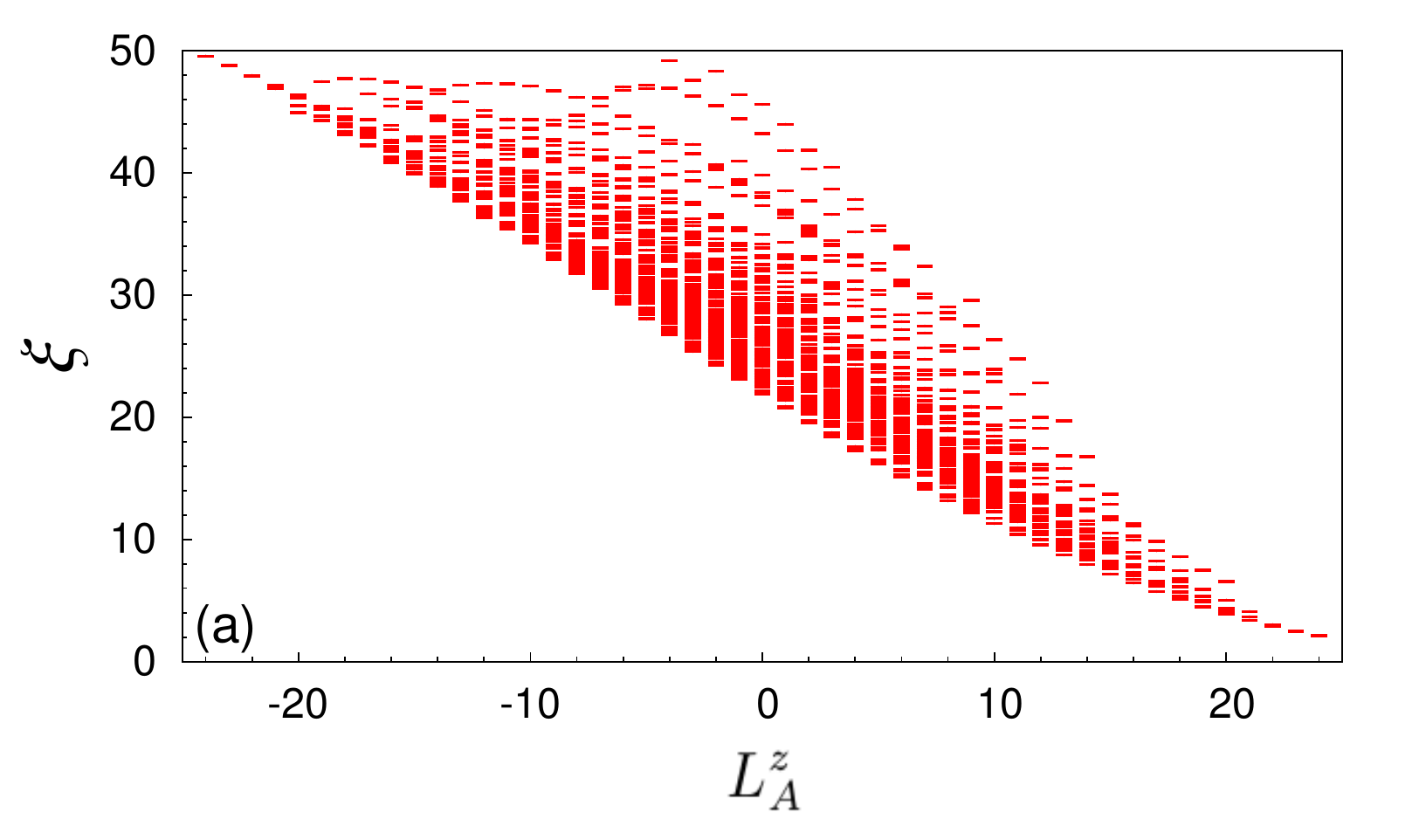}\\
\includegraphics[width=\linewidth]{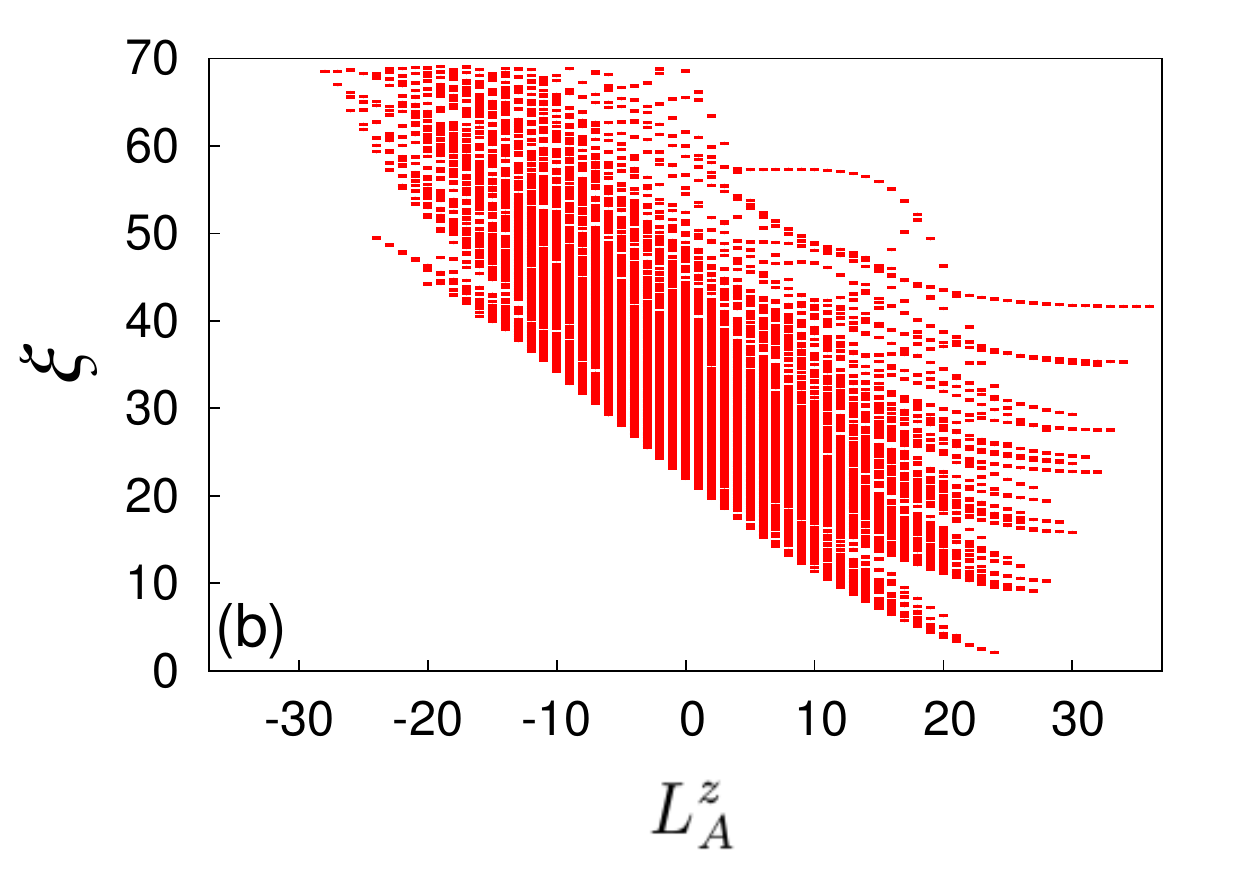}
\caption{The RES of (a) the $\nu = 1/3$ Laughlin state and (b) the $\nu=1/3$ Coulomb interaction ground state, both for $N=8$ particles with $N_A=4$, on a sphere partitioned into two hemispheres ($\Theta = \pi/2$).}
\label{reslaughlin}
\end{figure}

\begin{figure}[t!]
\includegraphics[width = \linewidth]{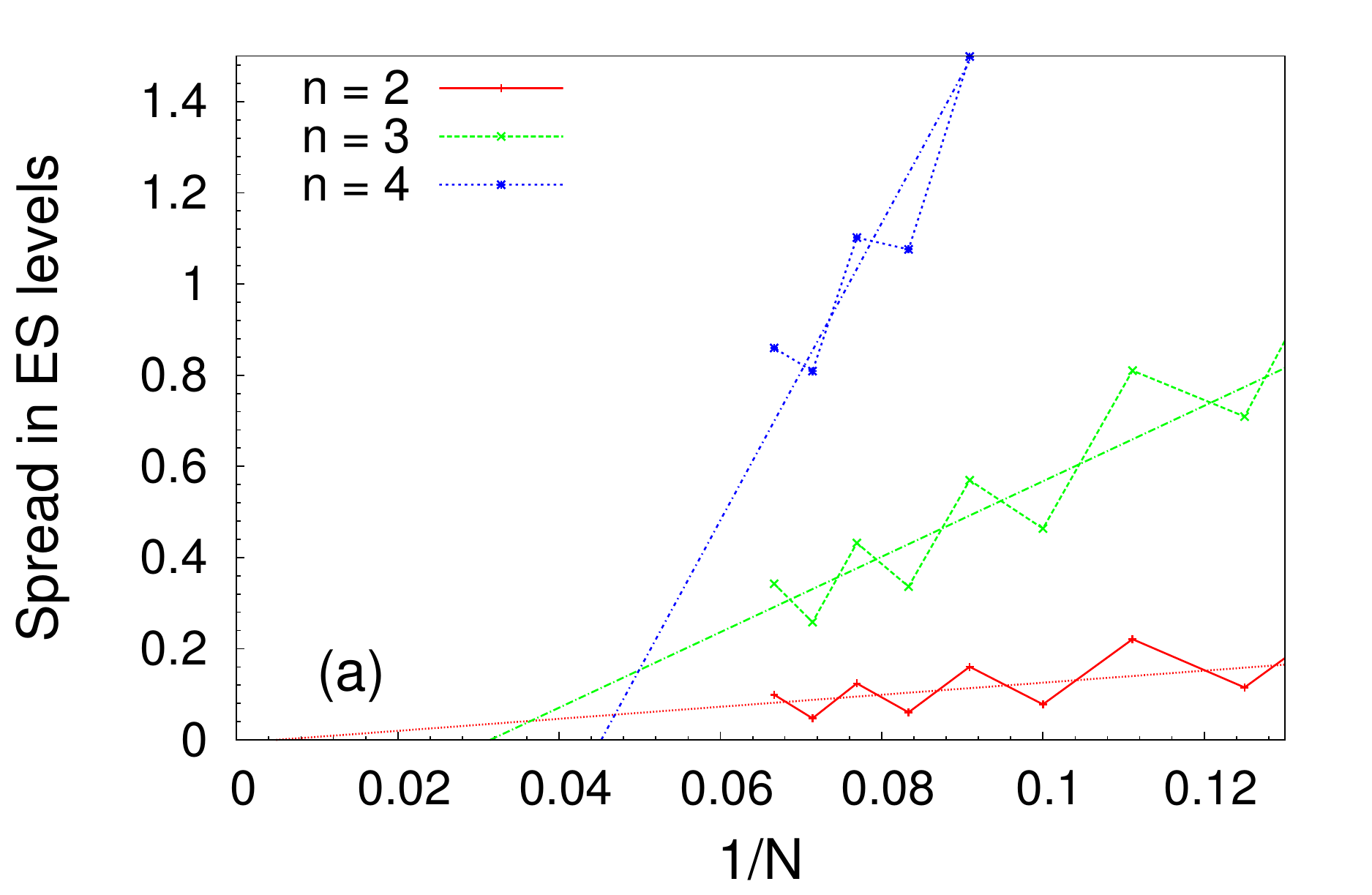}
\includegraphics[width = \linewidth]{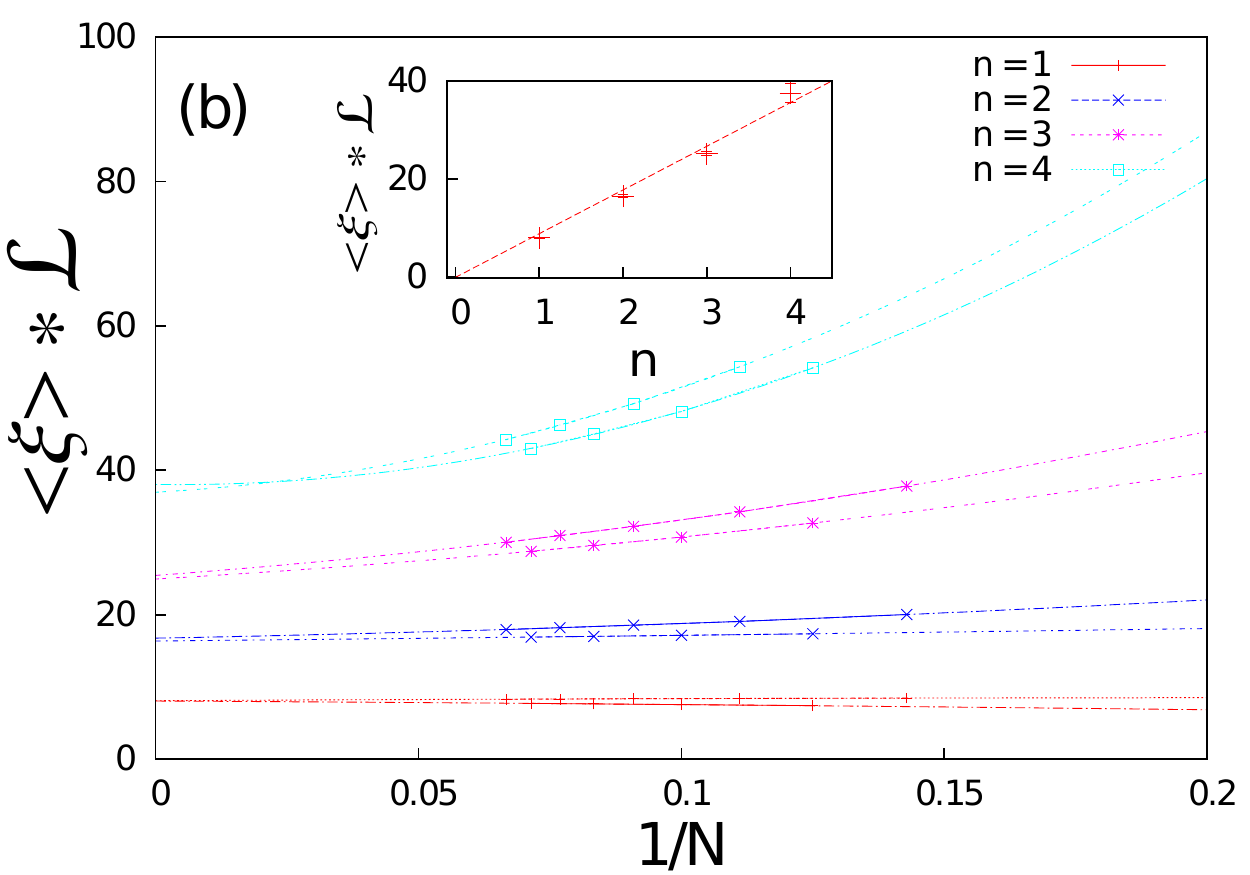}
\caption{(a) The spread in ES levels vs 1/$N$ for different $L_A^z$ sectors, using $\Theta = \pi/2$. An even-odd particle number effect is observed, and the spread of the levels vanishes in the thermodynamic limit. (b) The zero-spread ES levels extrapolated at the thermodynamic limit. Here $n=\max [L_A^z ] -L_A^z$. The even-odd effect is just a consequence of $N_A$ being the integer part of $N/2$. The velocity of the edge mode is $v = 1.41(5)$. Such value would be compatible with a rescaling of the $\nu=1$ edge mode velocity with a factor $1/\sqrt{3}$.}
\label{laughlinedgedispersion}
\end{figure}

We next consider the $\nu =1/3$ Laughlin state (Fig.~\ref{reslaughlin}a). As expected, the counting of the RES levels is given by the counting of the quasiholes states of $N_A$ particles in $N_\phi +1$ orbitals at momentum $L_A^z$. The RES of this state is chiral and exhibits more pronounced non-linearities than the RES of the $\nu=1$ IQH state, but should becoming increasingly linear as the thermodynamical limit is approached, as shown in Fig.~\ref{laughlinedgedispersion}. Notice that similar results have already been obtained for the OES in Ref.~\onlinecite{thomale-10pr180502}.

\subsection{$\nu=2/3$ Fractional Quantum Hall}

\begin{figure}[t!]
\includegraphics[width = \linewidth]{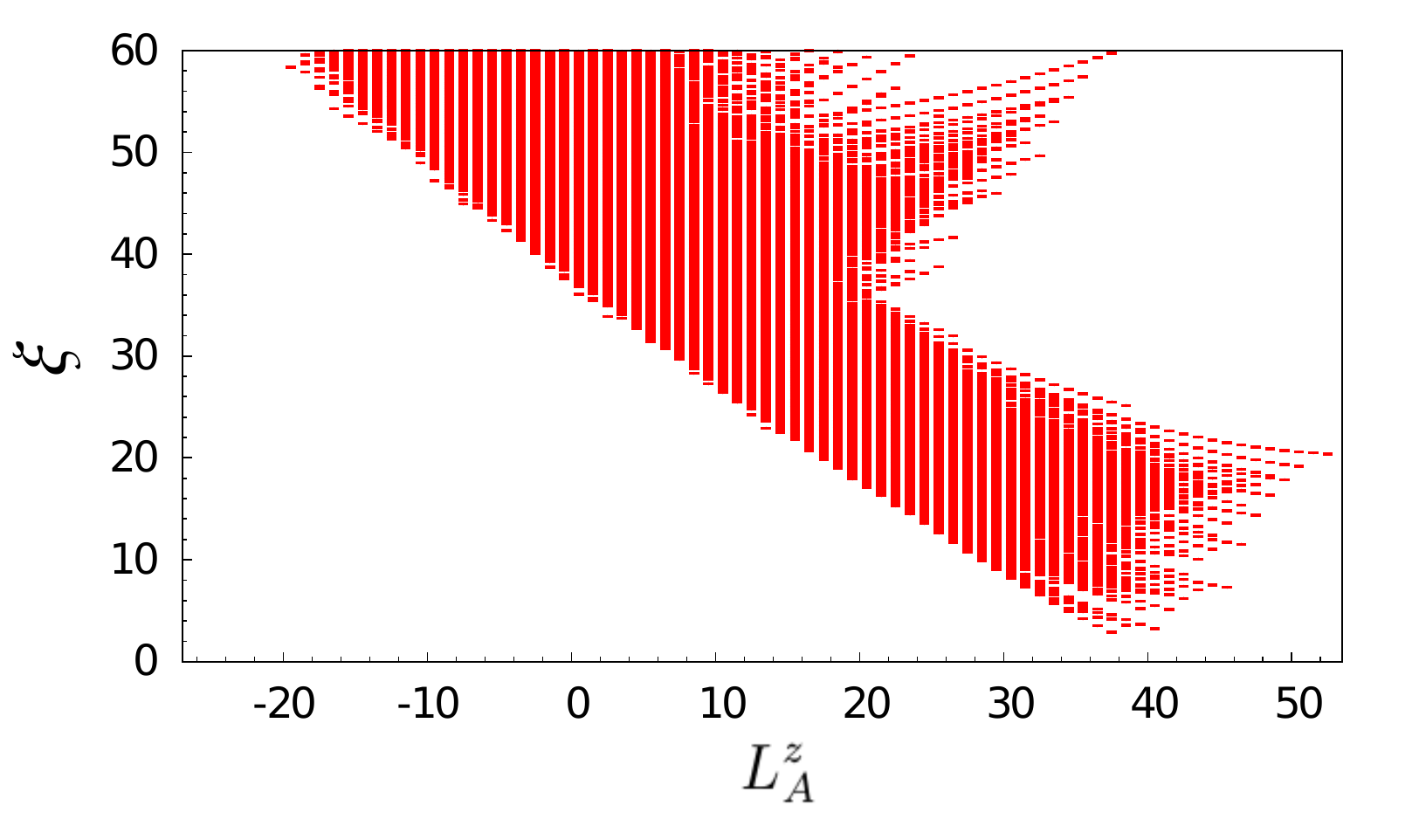}
\caption{The RES of the $\nu = 2/3$ particle-hole conjugate of the Laughlin state for $N = 14$, $N_A = 7$, and $\Theta = \pi/2$.}
\label{reslaughlinphconjugate}
\end{figure}

The RES is also able to exhibit the full edge mode structure of particle-hole conjugate states, which combine chiral and anti-chiral edge modes. In Fig.~\ref{reslaughlinphconjugate}, we plot the RES of the $\nu=2/3$ particle-hole conjugate of the Laughlin state. Unlike the OES (see Fig.~\ref{particleholelaughlin}), the structure of the RES reveals the presence of both a chiral and an anti-chiral modes. Starting from the maximal value of $L_A^z$, the entanglement levels first have an overall anti-chiral shape with negative velocity, followed by an overall chiral shape with positive velocity. In finite size samples, the two modes cover different total ranges of  $L_A^z$ (see Fig.~\ref{reslaughlinphconjugate}) due to the different compactification radii of the U(1) CFTs in this model. At intermediate values of $L_A^z$, the two modes combine in finite size, and their counting matches that of two mixed chiral and anti-chiral branches. Even though the RES can detect the presence of counter-propagating modes, it is not reliable in pining down the topological order of particle-hole conjugate FQH states. For example, given the RES of Fig.~\ref{reslaughlinphconjugate}, and no other information, we could immediately see that the state has counter-propagating edge modes, but could not infer that the state is the particle-hole conjugate of the $\nu=1/3$ Laughlin state. This is due to the fact the spectrum is filled with levels obtained by exciting some number of chiral and anti-chiral modes. However, having earned the information that the state has counter-propagating modes, we could now plot the OES of the $\nu=2/3$ state, which would give us the mirror-image of the $\nu=1/3$ Laughlin OES (see Fig.~\ref{particleholelaughlin}). The two pieces of information allow one to identify the state as the particle-hole conjugate of the $\nu=1/3$ Laughlin state.

\subsection{$\nu=1$ Bosonic Moore-Read State}

We have also computed the RES for non-Abelian quantum Hall states. While the study of the $\nu=5/2$ is highly relevant for FQH in experimental systems, we will focus on the simpler analog that occurs for $\nu=1$ bosonic systems. The fermionic $\nu=5/2$ state involves the second ($n=1$) Landau level, which would require to use the corresponding one-body wavefunctions, instead of those displayed in Eq.~(\ref{onebodywf}). The $\nu=1$ bosonic case is still a purely lowest Landau level problem. This system is also relevant for rapidly rotating ultra-cold atomic gases~\cite{cooper-08ap539}. For such systems, the physics is similar to that of FQH systems, except with bosons rather than electrons, and  replacing the Coulomb interaction with the short range (two-body) $\delta$-function interaction. There are several numerical studies that indicate this $\nu=1$ bosonic system is accurately described by the bosonic MR state~\cite{cooper-PhysRevLett.87.120405,regnault-PhysRevLett.91.030402,chang-PhysRevA.72.013611}.

The RES for the bosonic MR is shown in Fig.~\ref{resmooreread}a. As expected, its level counting matches that of the CFT edge mode counting. The corresponding RES for the delta-function interaction is depicted in Fig.~\ref{resmooreread}b. There is a clear entanglement gap and the low-lying branch has counting, shape, and values almost identical to those of the bosonic MR state. The structure of the spectrum above the gap will be analyzed in the following section.

\section{Higher Levels in the Entanglement Spectra}
\label{sec:higher_levels}

Similar to the situation for the OES, the RES of the Coulomb interaction ground state at $\nu=1/3$ (shown in Fig.~\ref{reslaughlin}b) exhibits the same low-lying structure as the RES for the $\nu=1/3$ Laughlin state. Additional branches are also clearly observed in the higher ES levels. These branches form what was thought to be the ``non-universal'' part of the Coulomb ES and are absent in the Laughlin model state ES. Similar to Ref.~\onlinecite{2011arXiv1105.5907S} for OES, we find that the RES branches of the Coulomb interaction ground-state are organized in a hierarchical structure that mimics the excitation-energy structure of the model pseudopotential Hamiltonian that has its ground state given by the Laughlin state. These structures can be accurately modeled using quasihole-quasielectron excitation wavefunctions. The study of Ref.~\onlinecite{2011arXiv1105.5907S} can be repeated here for the RES, with similar results.

\begin{figure}[t!]
\includegraphics[width=\linewidth]{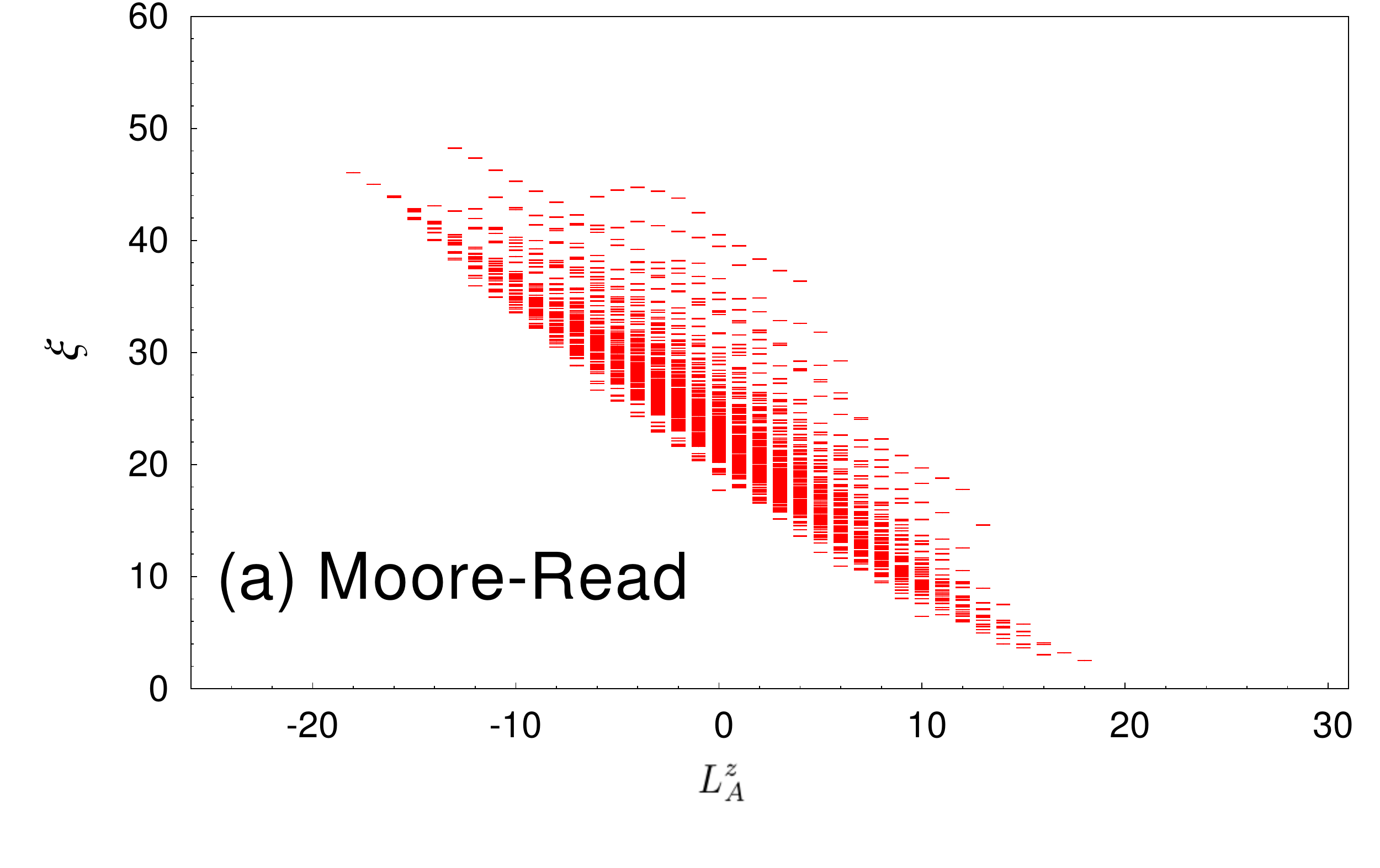}\\
\includegraphics[width=\linewidth]{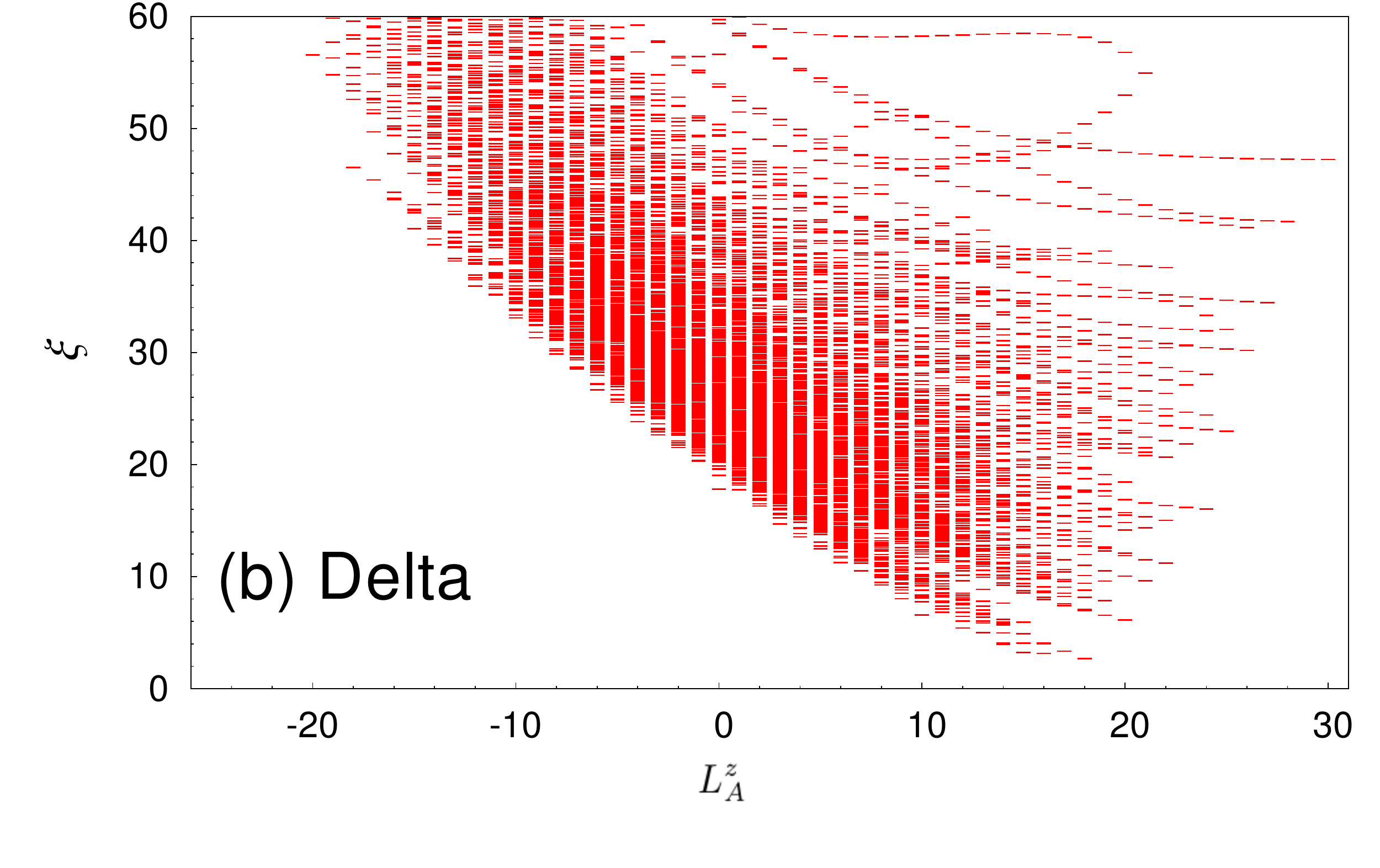}\\
\includegraphics[width=\linewidth]{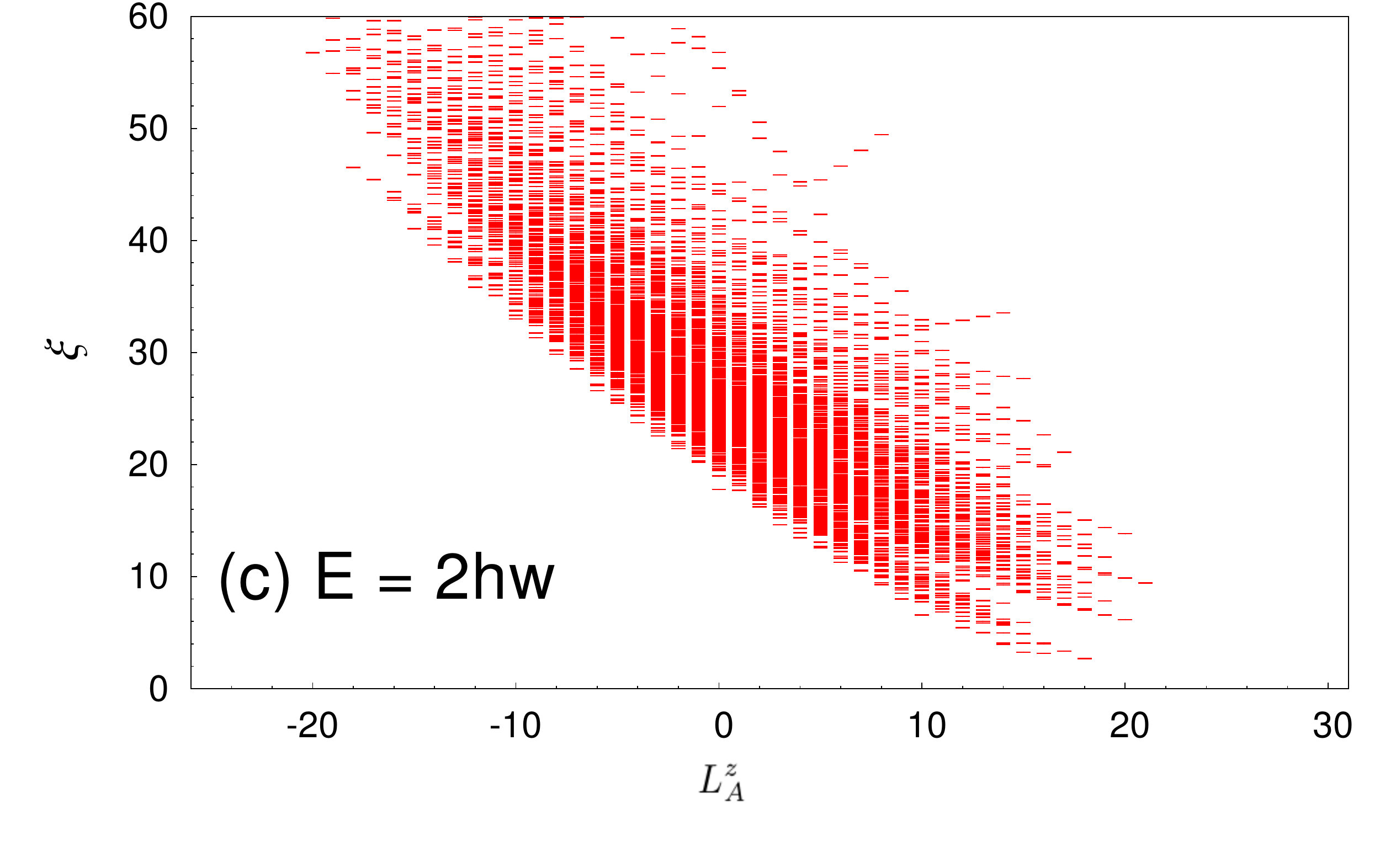}
\caption{The RES of (a) the $\nu = 1$ bosonic MR state, (b) the $\delta$-interaction ground state, and (c) an approximate state of $\delta$-interaction ground state involving up to $2$-quasihole-$2$-quasielecton excitations for $N=12$ particles with $N_A=6$. The sphere was partitioned into two hemispheres ($\Theta = \pi/2$).}
\label{resmooreread}
\end{figure}

The non-Abelian states also exhibits a series of higher energy branches that mimic the excitation-energy structure of the model pseudopotential Hamiltonian that has its ground state given by the MR state. These structures can be accurately modeled using the non-Abelian quasihole-quasielectron wavefunctions described in Ref.~\onlinecite{Rodriguez:2011aa} (these excitations can be obtained by taking two layers of Laughlin $\nu=1/2$ states, forming quasielectron-quasihole excitations in these layers, then symmetrizing over the electron coordinates between the layers), and the results are presented in Fig.~\ref{resmooreread}.

\section{Real Space Entanglement Spectrum of Quasihole Excitations}
\label{sec:quasiholes}

\begin{figure}[t!]
\includegraphics[width=\linewidth]{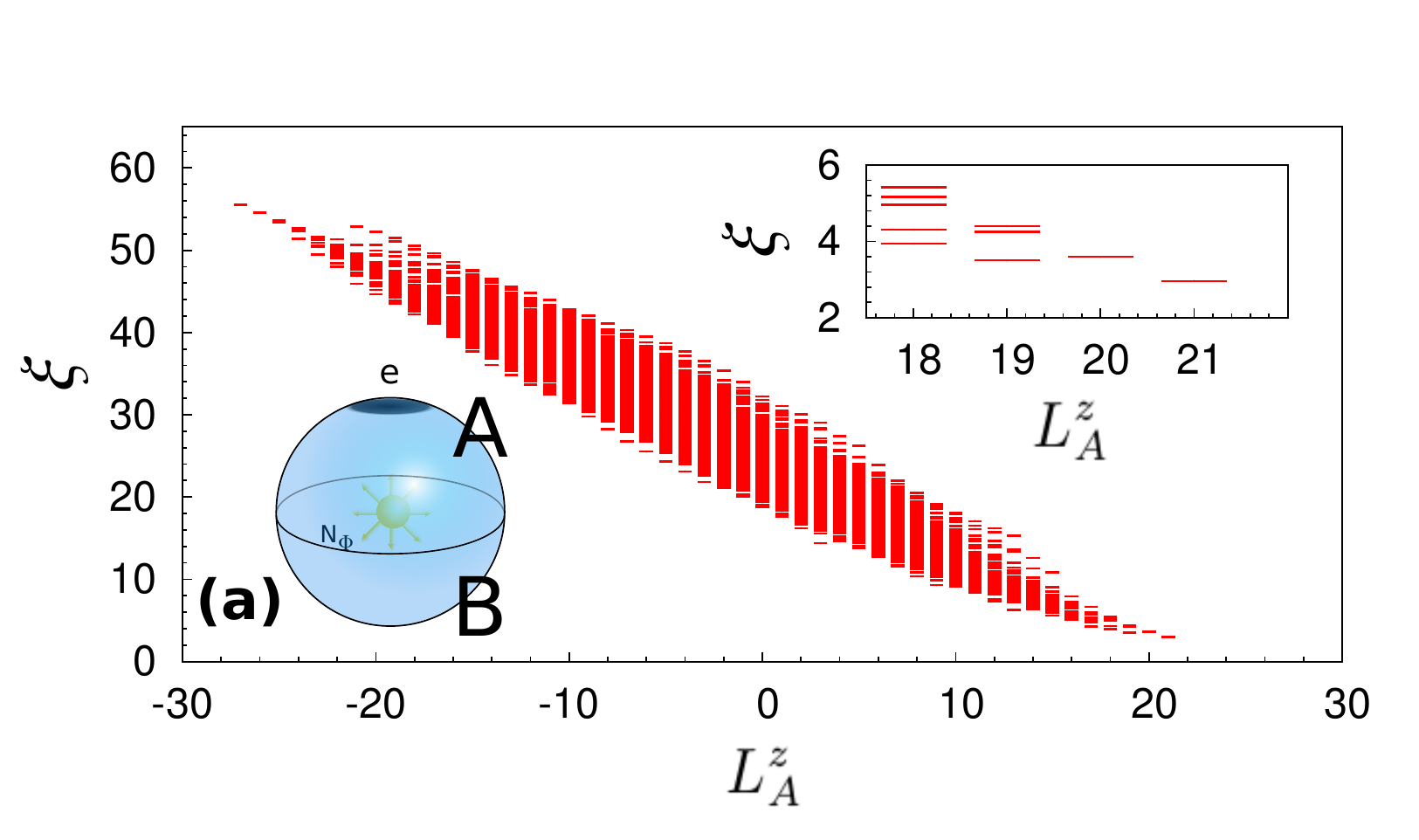}\\
\includegraphics[width=\linewidth]{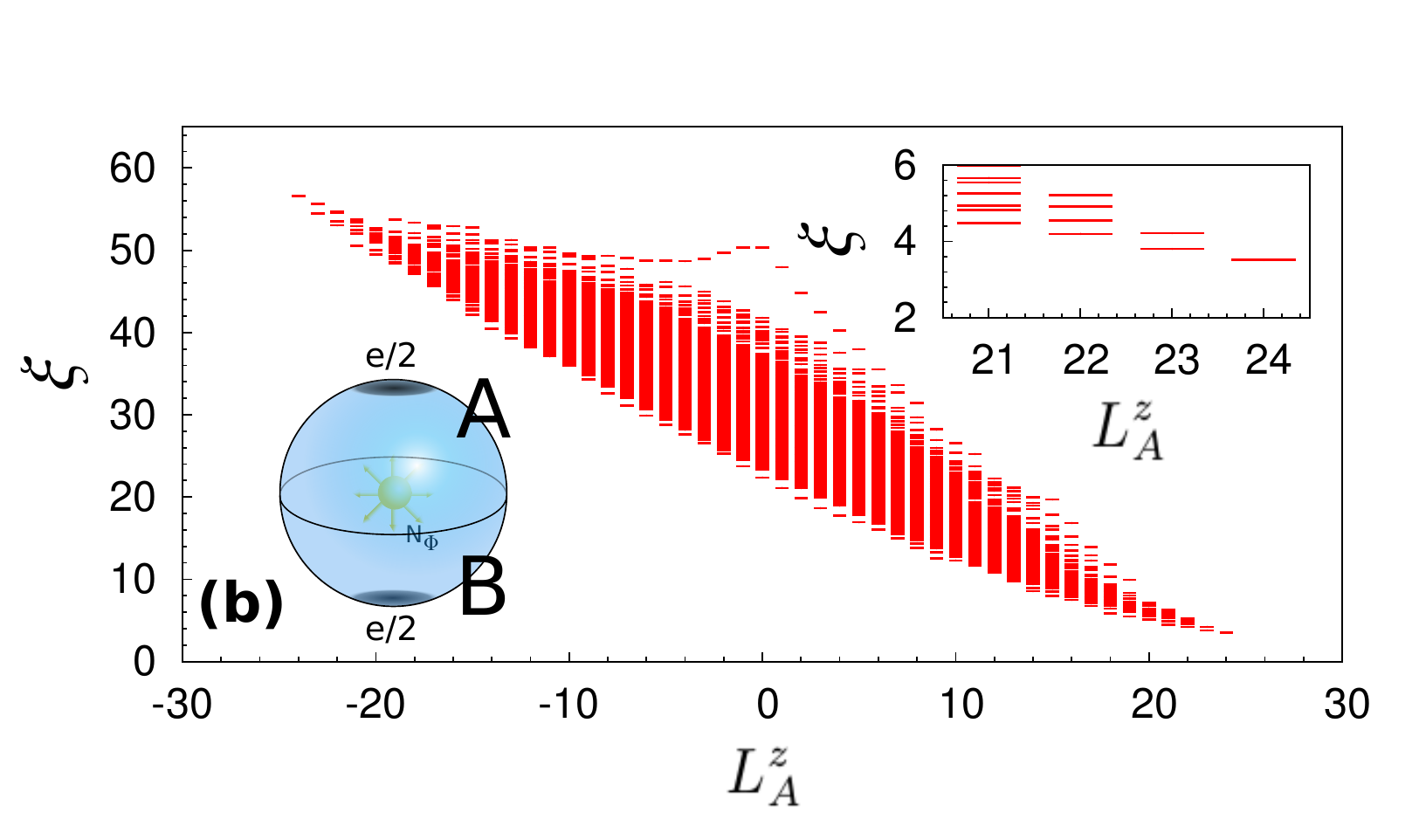}\\
\caption{(a) The RES for the ($\nu=1$ bosonic) MR state with $N=14$ and one Abelian quasihole at the north pole. The wavefunction for this configuration is the Jack polynomial with root partition $02020202020202$. (b) The RES for the MR state with $N=14$ and one non-Abelian quasihole at the north pole and one non-Abelian quasihole at the south pole. The wavefunction for this configuration is the Jack polynomial with root partition $11111111111111$. For both of these, the sphere was partitioned into two hemispheres ($\Theta = \pi/2$) and the particle sector is $N_A=6$.}
\label{resquasihole}
\end{figure}

We analyze the change in the RES for the MR state as a non-Abelian quasihole is taken from the north pole to the south pole across the real-space partition boundary. The evolution of the OES under this operation was analyzed in Ref.~\onlinecite{papic-PhysRevLett.106.056801}. The RES transformation is similar to that of the OES. Partitioning the system in real space at $\Theta = \pi/2$, for the state with two quasiholes at the north pole in the vacuum ($I$) channel (equivalently, one Abelian quasihole sits at the north pole), we see an entanglement spectrum that mimics that of the pure MR ground-state, with the counting $1,1,3...$ (see Fig.~\ref{resquasihole}). Taking one of these non-Abelian quasiholes across the sphere to the south pole brings us to the state with one quasihole at the north pole and one at the south pole. The counting in the RES is now $1,2,4...$ which is the that of the MR state with a $\sigma$ charge on the edge. The change in the counting occurs when the particle crosses the partition boundary.

\section{Real Space Entanglement Spectrum on the Annulus}
\label{sec:annulus}

\begin{figure}[t!]
\includegraphics[width=\linewidth]{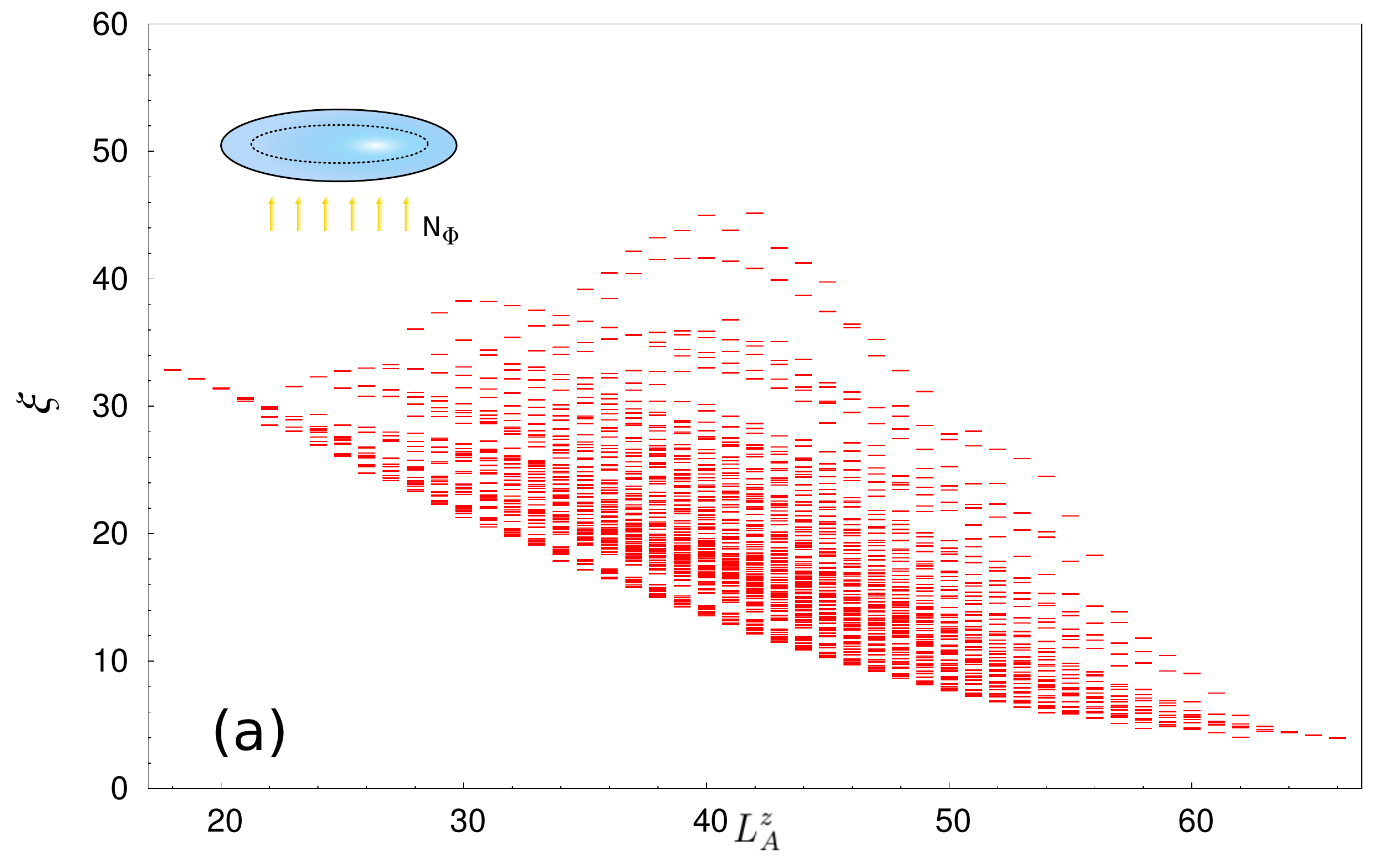}\\
\includegraphics[width=\linewidth]{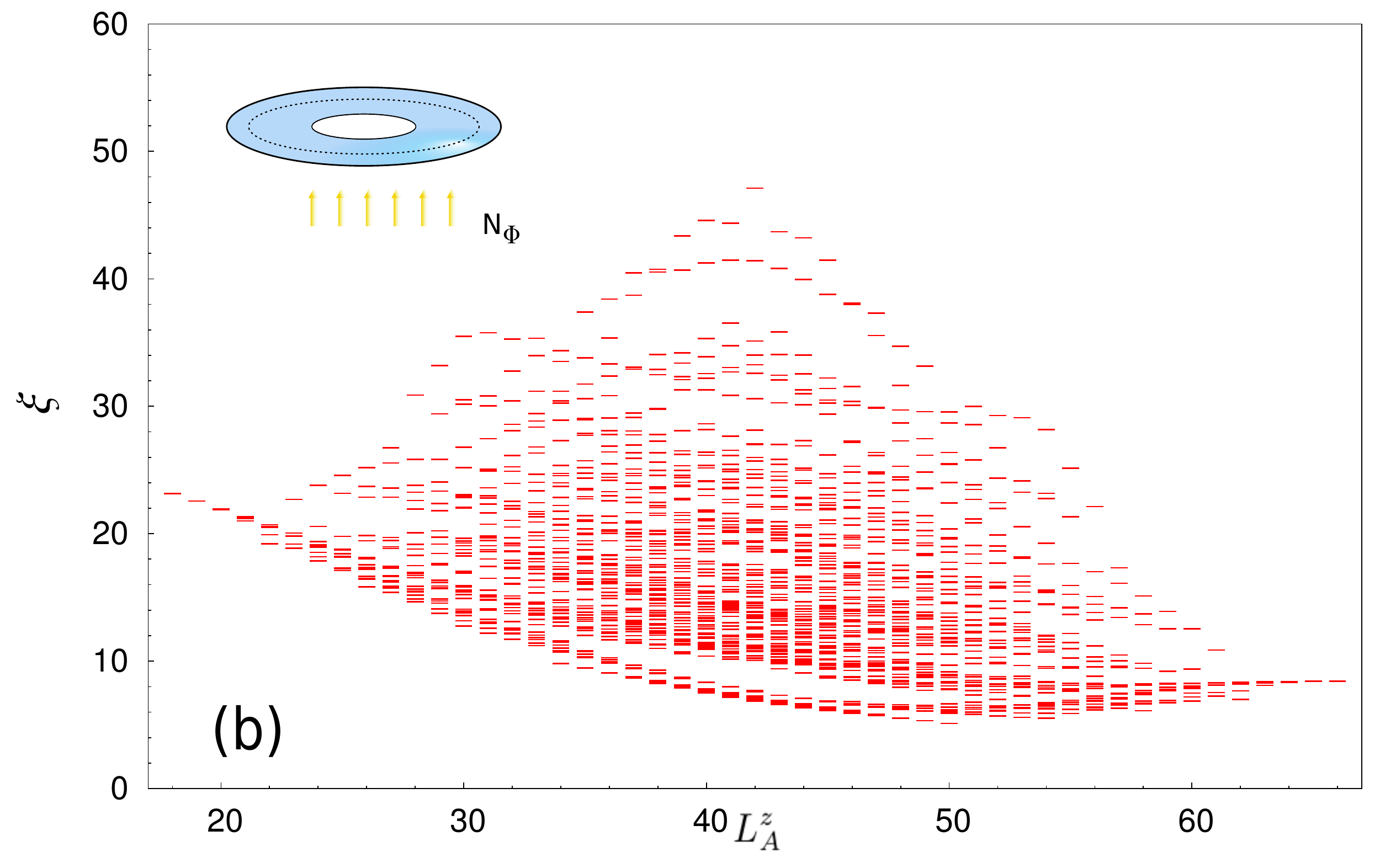}\\
\includegraphics[width=\linewidth]{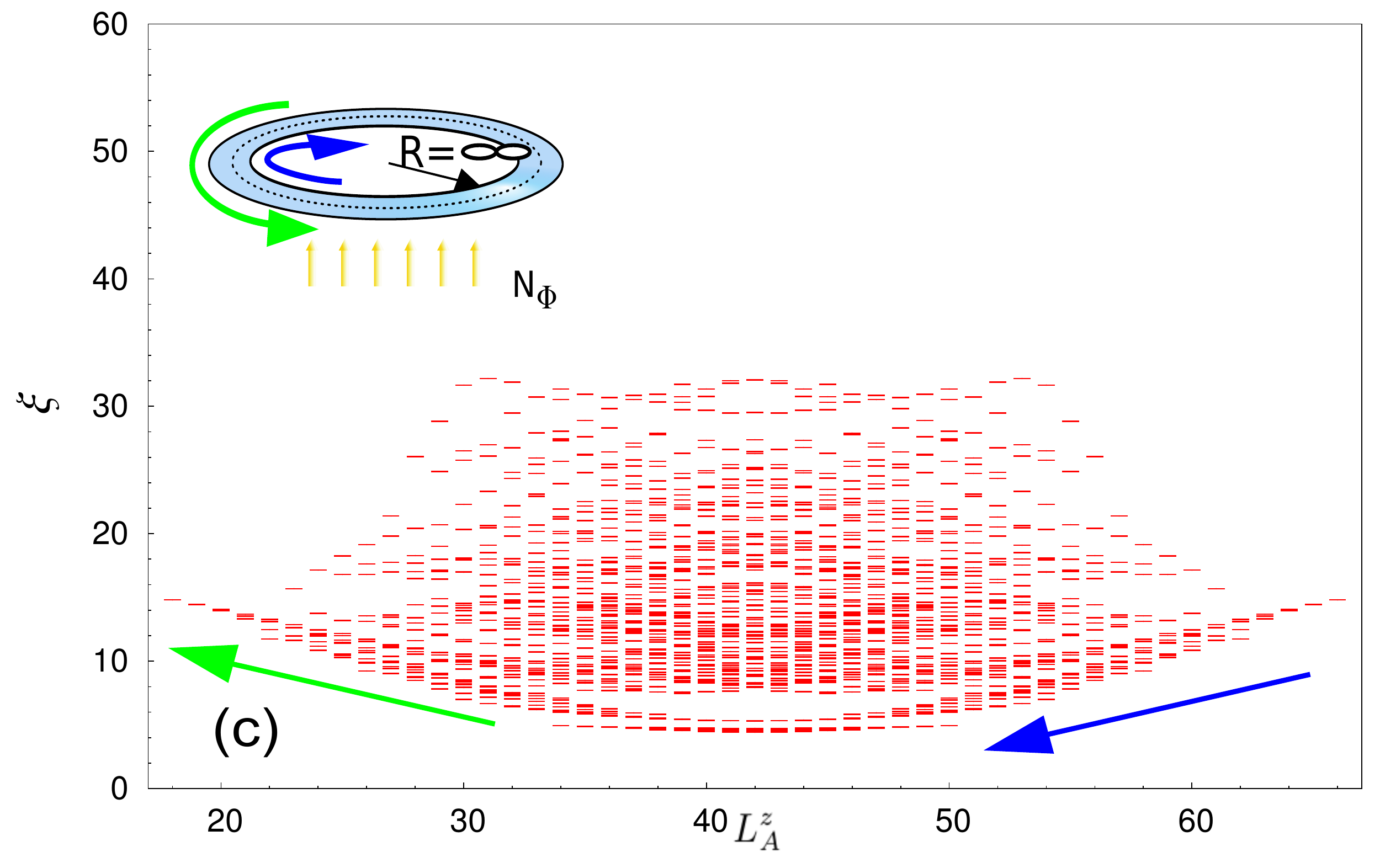}\\
\caption{Evolution of the RES with the changing of sample geometry from the disk to the thin annulus. (a) The RES for the Laughlin state with $N=8$ for the disk geometry, with a partition made at radius $R= \sqrt{N_{\phi}}$. The RES exhibits the usual chiral structure. (b) The RES for the Laughlin state on a thick annulus, obtained by inserting $20$ quasiholes at the origin. Note that the spectrum has flattened from the case of the disk. (c) The RES for the Laughlin state on the thin annulus, obtained by inserting a infinite number of quasiholes at the origin. The spectrum has a flat portion, sandwiched between a chiral and anti-chiral region (see the blue and green arrows) of states corresponding to the edge modes on the two sides of the annulus.}
\label{annulusRES}
\end{figure}

We now analyze the evolution of the RES when the geometry of the sample is altered from the disk to the thin annulus. The annulus represents the geometry of the so-called ``conformal limit''~\cite{thomale-10pr180502} for the OES, in which one notices a complete separation of the low-lying model-state universal ES levels from the high-lying non-universal levels, for the Coulomb ground states of bosons at $\nu= 1/2$ and $1$ and fermions at $\nu=1/3$ and $5/2$. We start from the filled disk and then proceed to insert a large number of quasiholes in order to reach the thin annulus limit. For a state of fixed number of particles (we work with $N=8$), this pushes the FQH liquid into an annulus of thickness $\Delta R = N/(2 \pi R \mu)$. For $R\rightarrow \infty$, the radial spatial extent of the annulus vanishes, the partition boundary is close to both edges of the annulus, and the entanglement spectrum is influenced by the presence of the annulus edges. This gives rise to the chiral and antichiral spectra in Fig.~\ref{annulusRES}.

\section{Entanglement Entropy}
\label{sec:entropy}

\begin{figure}[t!]
\includegraphics[width=\linewidth]{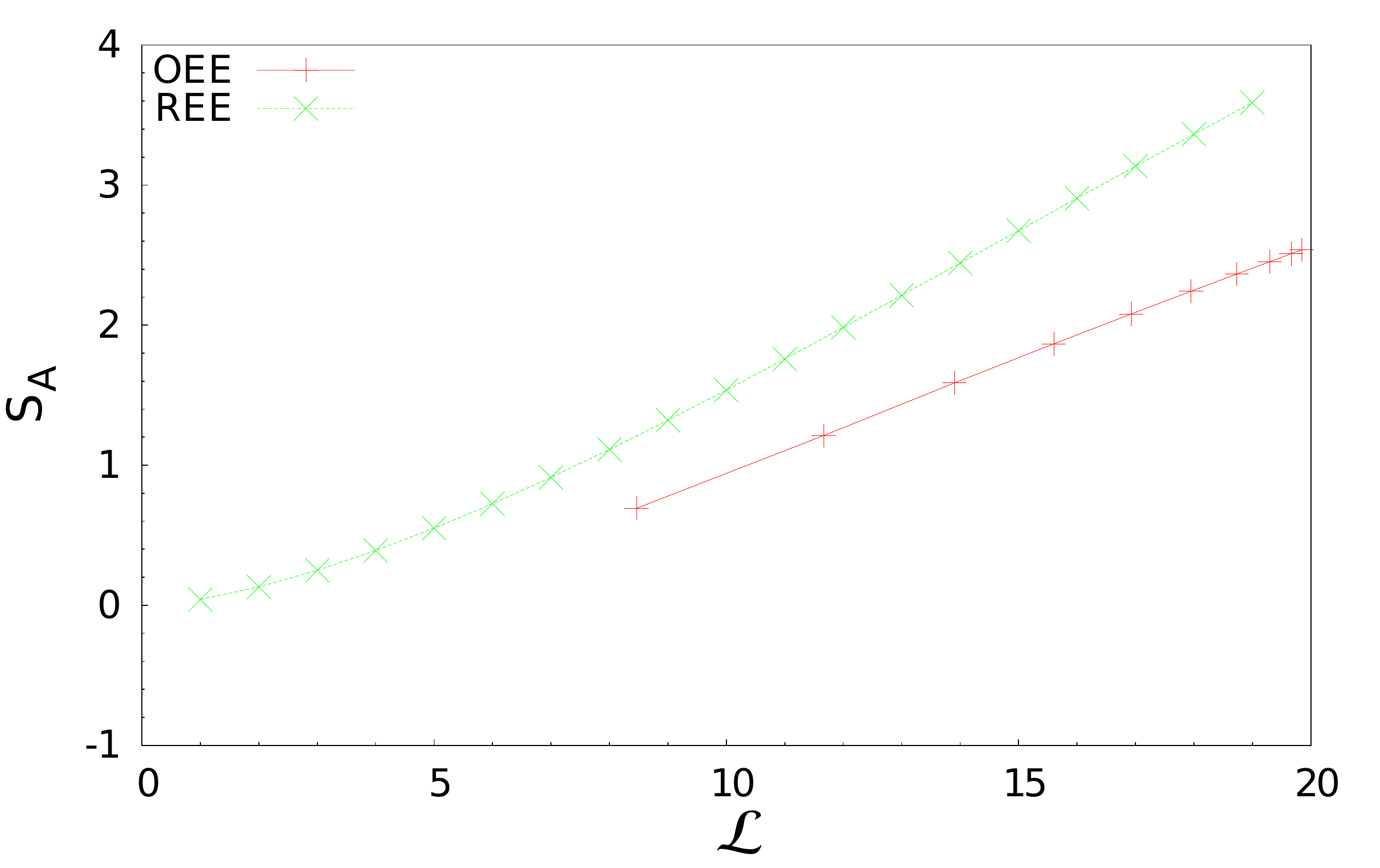}
\caption{Comparison between the entanglement entropy using the real-space partition and the orbital partition for the $\nu=1/2$ bosonic Laughlin state with $N=11$ particles.}
\label{compare_ree_oee}
\end{figure}

Using the real-space partition, we have computed the entanglement entropies for different model states in order to numerically verify the perimeter law and extract the value of the topological entanglement entropy. We compared the entanglement entropy obtained using a real-space partition with that obtained using an orbital partition, for the bosonic $\nu=1/2$ Laughlin state. As can be seen in Fig.~\ref{compare_ree_oee}, the difference between these can be quite substantial.

\begin{figure}[t!]
\includegraphics[width=\linewidth]{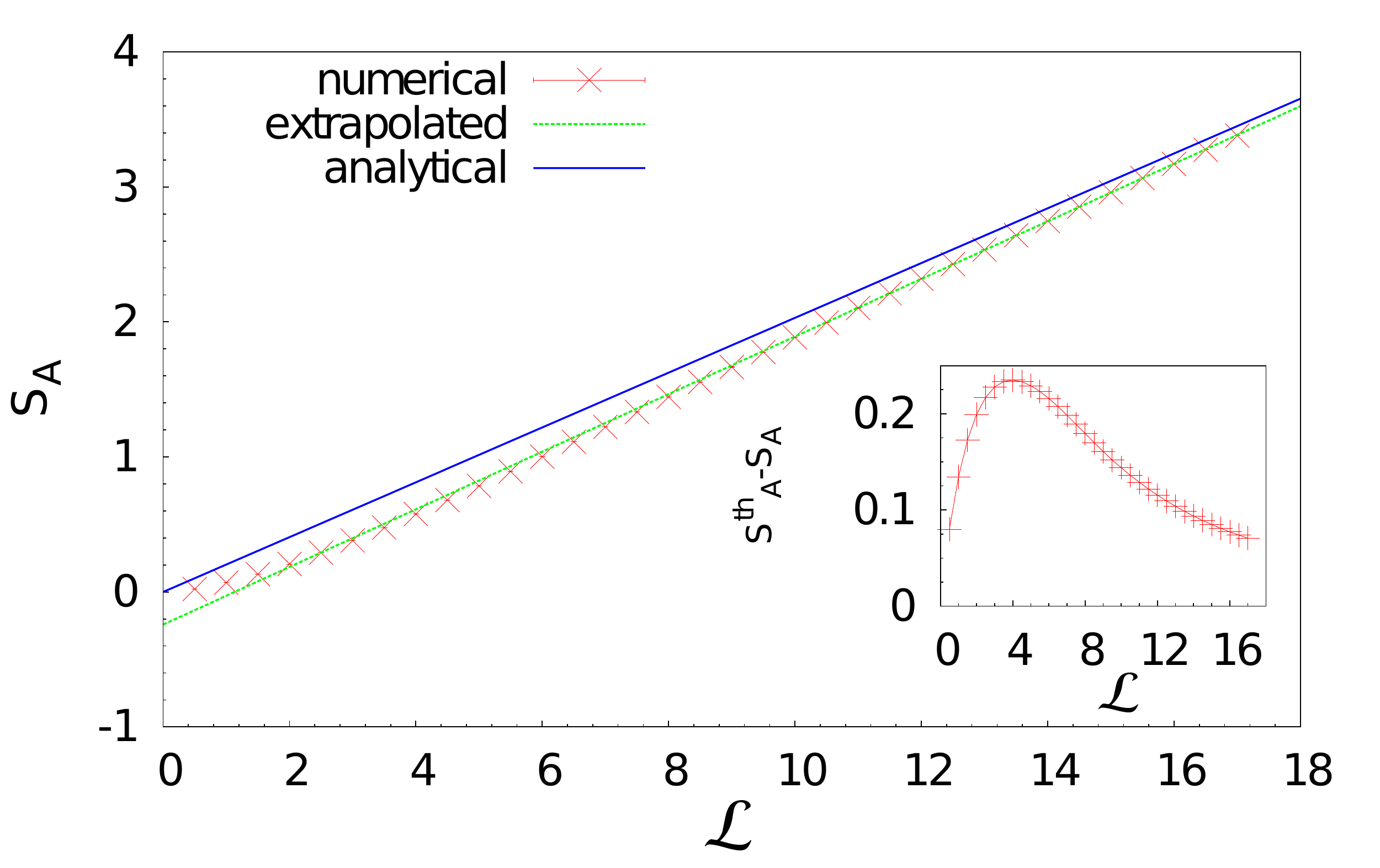}
\caption{Entanglement entropy as a function of the length $\mathcal{L}$ of the boundary between the two subspaces. The red $X$s are the (thermodynamic limit extrapolation of the) entanglement entropies for a given boundary length. The dotted green line is a linear fit of the entanglement entropies for $\mathcal{L}>10$. The solid blue line is the analytical result from Ref.~\onlinecite{PhysRevB.80.153303} that is valid in the $\mathcal{L}$ large limit. In the inset, we show the difference between the analytical and numerical values.}
\label{eeslater}
\end{figure}

We have also attempted to numerically verify the formula of Ref.~\onlinecite{PhysRevB.80.153303} for the entanglement entropy of the $\nu = 1$ IQH state in the thermodynamical limit: $S_A \simeq c_v \mathcal{L} -\gamma$ with a geometry-dependent $c_v = 0.203$ and $\gamma = 0$. For various boundary lengths $\mathcal{L}$ we compute the entanglement entropy for up to $N = 20$ particles, then calculate the thermodynamic value using a polynomial fit in $1/N$; this gives the thermodynamic limit of the entanglement entropy for a boundary of length $\mathcal{L}$. The estimated values and their errors are plotted in Fig.~\ref{eeslater} along with a linear fit, with only $\mathcal{L}>10$ values taken into account, of these data and the theoretical prediction for the $\mathcal{L} \gg 1$ regime. The linear fit of our data gives $\gamma = 0.241(5)$, which is not compatible with the expected value $\gamma =0$. However, the difference between our data and the analytical values vanishes for large $\mathcal{L}$ (see inset of Fig.~\ref{eeslater}). This indicates that, even for this very simple and ideal case, we are not able to reliably extract the value of the topological entanglement entropy without prior knowledge of its value (which would allow informed, but biased decisions on what data to keep) for the attainable system sizes in numerical studies. This casts doubt on prior numerical studies that claim to have extracted values of $\gamma$ that match the theoretical predictions.

\section{Conclusions}

In this paper we have analyzed the entanglement spectrum for a sharp, real-space partition of the space in which a FQH liquid exists. We have showed that the RES is intimately related (through diagonal transformations involving geometry dependent normalization factors)  to the PES and that, in particular, the two entanglement spectra have identical level counting. This counting is bounded by the number of CFT edge mode levels of the particular model state analyzed. It has been proven that this bound is saturated in the thermodynamic limit for these states and also at finite sizes for some of these states. We expect that the bound is saturated \emph{at any finite size} for all model states described by a CFT, and have numerically confirmed this for all sizes examined. We have also shown that the RES offers complementary information to the OES. In particular, the RES exhibits the expected (gapless) CFT edge spectra, even for the $\nu=1$ filled Landau level (IQH) state, and also reveals the presence of counter-propagating edge modes in particle-hole conjugate states, both of which fail to show up in the OES. For Coulomb states at fractional filling, the RES exhibits an entanglement gap and a series of higher entanglement energy branches whose level counting is in one-to one correspondence with that of the hierarchical quasihole-quasielectron excitations of the model state. This is similar to the information given by the OES; however, in the latter case, a full entanglement gap completely separating the low entanglement energy, universal levels from the spurious level, can sometimes exist by going to the conformal limit. In the case of the RES an entanglement gap exists only for the very few first levels of the spectrum. We have also analyzed the behavior of the RES as a non-Abelian quasihole is taken from the north pole to the south pole, passing across the partition boundary. The RES exhibits a change in its counting as the quasihole crosses the boundary, corresponding to the change between different topological sectors of the CFT (i.e. different topological charge on the boundary). Future research will focus on the real-space partitioning on the torus, on which the OES was investigated in Ref.~\onlinecite{lauchli:2010} and the PES in Ref.~\onlinecite{PhysRevLett.106.100405}.

\emph{Note added}: During the completion of this manuscript, we learned of similar unpublished works~\cite{Dubail-unpublished,Rodriguez-unpublished}.

\begin{acknowledgments}
BAB thanks D.~Haldane, M.~Hermanns, and T.~Hughes for fruitful discussions. PB, AC, NR, and AS thank J.~Dubail, D.~Haldane, M.~Hermanns, T.~Hughes, N.~Read, E.~Rezayi, I.~Rodriguez, S.~Simon, and J.~Slingerland for useful discussions. BAB was supported by Princeton Startup Funds, NSF CAREER DMR-095242, ONR - N00014-11-1-0635, Darpa - N66001-11-1-4110 and David and Lucile Packard Foundation. BAB, PB, and NR acknowledge the CFT, Topology and Information Workshop at the Institut Henri Poincar\'e, where this work was completed. BAB, AC, and NR acknowledge the support and hospitality of Microsoft Station Q. PB acknowledges the hospitality and support of ENS-Paris, the Institut Henri Poincar\'e, and the Aspen Center for Physics under NSF grant No. $1066293$.
\end{acknowledgments}


\end{document}